\documentstyle [fleqn,12pt] {article}
\newcommand{\cd}{\cdot}

\newcommand{\al}{\alpha}
\renewcommand{\b}{\beta}
\newcommand{\de}{\delta}
\newcommand{\De}{\Delta}
\newcommand{\ep}{\epsilon}
\newcommand{\ga}{\gamma}
\newcommand{\Ga}{\Gamma}

\newcommand{\la}{\lambda}
\newcommand{\Om}{\Omega}
\newcommand{\om}{\omega}
\newcommand{\si}{\sigma}

\newcommand{\th}{\theta}

\newcommand{\ra}{\rightarrow}

\newcommand{\lap}{\triangle}

\newcommand{\bm}[1]{\mbox{\boldmath $#1$}}

\newcommand{\be}{\begin{equation}}
\newcommand{\ee}{\end{equation}}
\newcommand{\bea}{\begin{eqnarray}}
\newcommand{\eea}{\end{eqnarray}}
\newcommand{\bean}{\begin{eqnarray*}}
\newcommand{\eean}{\end{eqnarray*}}
\newcommand{\dd}{\partial}

\begin{document}

\begin{titlepage}
\begin{center}
{\huge Large Scale Structure Formation with Global Topological Defects}
	\vspace{0.2cm}\\
{\Large A new Formalism and its Implementation by Numerical 
	Simulations}
    \vspace{0.2cm}  \\  {\bf \large Ruth Durrer and Zhi--Hong Zhou}
    \vspace{1cm} {\large
   \\ Universit\"at Z\"urich, Institut f\"ur Theoretische Physik,
      Winterthurerstrasse ~190, \\ CH-8057 Z\"urich, Switzerland} 
\end{center}
\vspace{2cm}
\begin{abstract}
We investigate cosmological structure formation seeded by topological
defects which may form during a phase transition in the early universe.
 First we  derive a partially new,  local and gauge 
invariant system of perturbation equations to treat microwave background
and dark matter fluctuations induced by topological defects or any 
other type of seeds. We then show that this system is   well
suited for   numerical analysis  of structure formation by
applying it to seeds induced by fluctuations of a global scalar field. 
Our numerical results cover a
 larger dynamical range than previous investigations and are
complementary to them since we use substantially different methods.
The resulting microwave background fluctuations are  compatible with 
older simulations. We also obtain a scale invariant spectrum of 
fluctuations with about the same amplitude. However, our dark matter 
results yield a smaller bias parameter compatible with $b\sim 2$ on
a scale of $20 Mpc$ in contrast to   previous 
work which yielded to large bias factors. Our conclusions are thus 
more positive. According to the aspects analyzed
 in this work, global topological defect induced fluctuations 
yield viable scenarios of structure formation and do better than 
standard CDM on large scales.
\end{abstract}
 \vspace{1cm}


\end{titlepage}

\section{Introduction}

The formation of cosmological structure in the universe, 
 inhomogeneities in the matter distribution like quasars at redshifts 
up to $z\sim 5$, galaxies, clusters, 
super clusters, voids and walls,  is an outstanding basically 
 unsolved problem within the standard model of cosmology. 

At first sight it seems obvious that small density enhancements can grow
sufficiently rapidly by gravitational instability. But global 
expansion of the universe and
radiation pressure counteract gravity, so that, e.g., in the case of
a radiation dominated, expanding universe no density inhomogeneities
can grow faster than logarithmically. Even in a universe dominated by 
pressure-less matter, cosmic dust, the
growth of density perturbations is strongly reduced by the expansion
of the universe.

On the other hand, we know that the universe was extremely homogeneous
and isotropic at early times. This follows from the isotropy of
the 3K Cosmic Microwave Background (CMB), which represents a relic of
the  plasma of baryons, electrons and radiation at times before protons 
and electrons combined to hydrogen. After a long series of upper bounds,
measurements with  the COsmic Background Explorer satellite  (COBE)
have finally established anisotropies in this radiation \cite{Sm} at 
the level of 
\[   \langle{\De T\over T}(\th)\rangle \sim 10^{-5} 
	~~~\mbox{ on  angular scales}~
		 7^o\le \th\le 90^o ~.\]
On smaller angular scales  the observational situation is 
at present somewhat confusing and contradictory \cite{WJNPW}, but many 
upper limits require 
$ \De T/T < 4\times 10^{-5} $ on all scales $\th <8^o$.  

All observations together clearly rule out the simplest model of a 
purely baryonic universe with density parameter $\Om \sim 0.1$ and 
adiabatic initial fluctuations (either the initial perturbations are too
large to satisfy CMB limits, or they are too small to develop into the
observed large scale structure).

The most conservative way out, where one just allows for non--adiabatic
initial perturbations (minimal isocurvature model), also faces severe 
difficulties \cite{PS,GSS,PD,HBS}. In
other models one assumes that  initial fluctuations are
created during an inflationary epoch, but that the matter content of 
the universe is dominated by  hot or  cold dark matter or a mixture
of both. Dark matter particles do not interact with photons other than
gravitationally  and thus induce perturbations in the CMB only via 
gravitation. In these models, inflation generically leads to $\Om =1$, 
while the baryonic density parameter is
only  $\Om_Bh^2 \sim 0.02$, compatible with nucleosynthesis constraints.
 With  one 
component of dark matter, these models  do not seem to 
agree with observations \cite{GSS,Os}, however, if a suitable mixture 
of hot and cold dark matter is adopted, the results from numerical 
simulations look quite promising \cite{Ho,XK,DSS}, although they might
have difficulties to account for the existence of clusters at a redshift
$z\sim 1$ \cite{Ba}.
 
In  these dark matter models  initial fluctuations are 
generated during an inflationary phase. Since all worked out models
of inflation face  difficulties (all of them have to invoke  fine 
tuning to obtain the correct amplitude of density inhomogeneities), we 
consider it very important to investigate yet another possibility: 
Density perturbations in the dark matter and baryons might have been 
triggered by seeds.
 Seeds are an inhomogeneously distributed form of energy which
makes up only a small fraction of the  total energy density of the 
universe. 
Particularly natural seeds are topological defects. They can form
during symmetry breaking phase transitions in the early universe
 \cite{Ki,CDTY}. Depending on the symmetry being gauged or global, the
corresponding defects are called local or global.

The fluctuation spectrum on large scales observed by COBE is
not very far from scale invariant \cite{Go}. This has been 
considered a great success for inflationary models which generically 
predict a scale
invariant fluctuation spectrum. However, as we shall see, also
models in which perturbations are seeded by global topological defects 
yield  scale invariant spectra of CMB fluctuations. To be specific, we 
shall mainly
 consider texture, $\pi_3$--defects which lead to event singularities
in four dimensional spacetime \cite{Tu,d95}.  Global defects are viable
candidates for structure formation, since the scalar field energy 
density, $\rho_S$, of global topological defects  scales like
 $\rho_S \propto 1/(at)^{2}$ (up to a logarithmic correction for global
strings) and thus always represents  the same fraction of the total 
energy density of the universe ($t$ is conformal time).
\begin{equation} \rho_S/\rho \sim 8\pi G\eta^2 \equiv 2\epsilon ~, 
\end{equation}
where $\eta$ determines  the symmetry breaking scale ({\sl see} Fig.~1). For
the background spacetime we assume a Friedmann--Lema\^{\i}tre universe 
with $\Omega=1$ dominated by cold dark matter (CDM). We choose 
conformal coordinates  such that
\[ ds^2 = a^2(-dt^2 + \delta_{ij}dx^idx^j)  ~. \]
Numerical analysis of CMB fluctuations from topological defects on large
scales  has been performed in \cite{BR,PST}; a spherically  symmetric 
approximation is discussed in \cite{DHZ}. Results for intermediate scales
angular are presented in \cite{CFGT}. All these investigations (except  
\cite{DHZ}) use linear cosmological perturbation theory in synchronous
gauge and (except \cite{PST}) take into account only scalar 
perturbations. 
Here we derive a fully gauge invariant and local system of 
perturbation equations. The (non--local) split into 
scalar, vector and tensor modes on  hyper surfaces of constant time 
is not performed. We  solve the equations numerically in a cold dark
matter (CDM) universe with global texture. In this paper, we 
detail the results outlined in a previous letter \cite{DZ}. Furthermore,
we present  explicit
derivations of the equations, a description of  our numerical 
methods and we briefly discuss some tests of our codes.  
Since there are no spurious gauge modes in our initial conditions, 
there is no danger that these may grow in time and some of the 
difficulties to choose correct initial conditions ({\sl see} e.g. \cite{PST}) 
are removed. However, as we shall discuss in Section~3, the results
do depend very sensitively on the choice of initial conditions.

Nevertheless, we should keep in mind, that we are investigating models 
of structure formation which rely on the particle physics and cosmology
at temperatures of $T\sim T_{GUT} \sim 10^{16}GeV$. An energy scale
about which we have no experimental evidence whatsoever. The physical
model adopted for our calculations should thus always be considered as a
toy model, of which we hope it captures the features relevant for 
structure formation of the 'realistic physics' at these energies.
Therefore, we  suggest, not to take the results serious much beyond
about a factor of two or so. On the other hand, our
models show that the particle physics at GUT scale may have left its
traces in the  distribution of matter and radiation in the present 
universe, yielding the exciting possibility to learn about the physics 
at the highest energies, smallest scales, by probing the  largest
structures of the universe.
\vspace{0.2cm}

We calculate the CMB anisotropies on angular scales
which are larger than the angle subtended by the horizon scale at
decoupling of matter and radiation, $\theta> \theta_d$. For $\Omega =1$
and $z_d \approx 1000$
\begin{equation} \theta_d = 1/\sqrt{z_d+1} \approx 0.03 \approx 2^o 
	~. \end{equation}
It is therefore sufficient to study the generation and evolution 
of microwave background fluctuations after recombination. During this
period,  photons stream freely, influenced only by cosmic 
gravitational redshift and by perturbations in the gravitational 
field (if the medium is not re-ionized). 

In Section~2 we derive a local and gauge invariant perturbation 
equation to calculate the CMB fluctuations. In Section~3, we
put together the full system of equations which has to be solved
to investigate gravitationally induced CMB fluctuations and the dark
matter perturbation spectrum in a model with global topological defects.
We  discuss  the choice of initial conditions and the numerical treatment
 of this system in Section~4. The next section is devoted to the
presentation and analysis of our numerical results. We end with 
conclusions in Section~6.
\vspace{0.2cm}

{\bf Notation:} We denote conformal time by $t$. Greek indices run from 
0 to 3, Latin indices run from 1 to 3. The metric signature is chosen 
$(- + + +)$. We set $\hbar=c=k_{Boltzmann}=1$ throughout.

\section{A Local and Gauge Invariant Treatment of the Perturbed Liouville
   	Equation}
Collision-less particles are described by their one particle 
distribution function which lives on the seven dimensional phase space 
\[ {\cal P}_m = \{ (x,p)\in T\!{\cal M}| g(x)(p,p)=-m^2\} ~. \]
Here $\cal M$ denotes the spacetime manifold and $T{\cal M}$ its 
tangent space. The fact that collision-less particles move on geodesics
translates to the Liouville equation for the one particle distribution
function, $f$. The Liouville equation reads \cite{Ste}
\be  X_g(f)=0    \label{L}~.\ee
In a tetrad basis $(e_\mu)_{\mu=0}^3$ of $\cal M$, the vector field
 $X_g$ on ${\cal P}_m$ is given 
by ({\sl see} e.g. \cite{Ste})
\be 
	X_g = (p^\mu e_\mu - \om^i_{\:\mu}(p)p^\mu{\dd \over \dd p^i}) 
  ~, \label{liou}	\ee
where $\om^\nu_{\:\mu}$ are the connection 1--forms of $({\cal M},g)$ 
in the basis $e^\mu$, and we have chosen the basis  
\[(e_\mu)_{\mu=0}^3~~ \mbox{ and }~~~
({\dd\over \dd p^i})_{i=1}^3 ~~\mbox{  on }~~~~  T{\cal P}_m~, 
	~~~~~~  p=p^\mu e_\mu~. \]
We now apply this general framework to the case of a perturbed Friedmann
universe. The metric of a perturbed Friedmann universe with density
parameter $\Om=1$ is given by $ds^2=g_{\mu\nu} dx^\mu dx^\nu$ with
\be 
 g_{\mu\nu} = a^2(\eta_{\mu\nu} + h_{\mu\nu}) = a^2 \tilde{g}_{\mu\nu}
  ~, \ee
where $(\eta_{\mu\nu}) = diag(-,+,+,+)$ is the flat Minkowski metric and
$(h_{\mu\nu})$ is a small perturbation, $|h_{\mu\nu}|\ll1$. We now use 
the fact that the motion of photons is conformally invariant: 

We  show that for massless particles and conformally related metrics,
\[g_{\mu\nu}= a^2\tilde{g}_{\mu\nu}~,\]
\be
	(X_gf)(x,p)=0 ~~~\mbox{ is equivalent to }~~~
	(X_{\tilde{g}}f)(x,ap)=0  ~. 
	\label{conform}		\ee
This is easily seen if we write $X_g$ in a coordinate basis:
\[ X_g = b^\mu\dd_\mu -\Ga_{\al\beta}^ib^\al b^\beta{\dd\over \dd b^i}
	~,\]
with 
\[ \Ga_{\al\beta}^i={1\over 2}g^{i\mu}(g_{\al\mu},_\beta + 
	g_{\beta\mu},_\al-g_{\al\beta},_\mu) ~. \]
The $b^\mu$ are the components of the momentum $p$ with respect to the
{\em coordinate} basis:
\[ p=p^\mu e_\mu = b^\mu\dd_\mu ~.\]
If $(e_\mu)$ is a tetrad with respect to $g$, then $\tilde{e}_\mu=ae_\mu$
is a tetrad basis for $\tilde{g}$. Therefore,  the coordinates of 
of $ap=ap^\mu\tilde{e}_\mu= a^2p^\mu e_\mu = a^2b^\mu\dd_\mu$ with 
respect to $\dd_\mu$ on 
$({\cal M},\tilde{g})$ are  given by $a^2b^\mu$.
In the coordinate basis thus our statement Eq.~(\ref{conform}) follows, if we
can show that
\be
	 (X_{\tilde{g}}f)(x^\mu,a^2b^i)=0 
	~~~\mbox{ iff }~~~  (X_gf)(x^\mu,b^i)=0 
	\label{conformco} \ee
Setting $v=ap=v^\mu\tilde{e}_\mu = w^\mu\dd_\mu$, we have $v^\mu=ap^\mu$
and $w^\mu=a^2b^\mu$. Using $p^2=0$, we obtain the following relation 
for the Christoffel symbols of $g$ and $\tilde{g}$:
\[ \Ga^i_{\al\beta}b^\al b^\beta=  \tilde{\Ga}^i_{\al\beta}b^\al b^\beta
  +{2a,_\al\over a}b^\al b^i  ~.\]
For this step it is crucial that the particles are massless! For massive
particles the statement is of course not true.
Inserting this result into the Liouville equation we find
\be
 a^2X_gf = w^\mu(\dd_\mu f|_b -2{a,_\mu\over a} b^i{\dd f\over \dd b^i})
  -\tilde{\Ga}_{\al\beta}^iw^\al w^\beta {\dd f\over \dd w^i} ~,
	\label{lstar} \ee
where $\dd_\mu f|_b$ denotes the derivative of $f$ w.r.t. $x^\mu$ at
constant $(b^i)$. Using
\[\dd_\mu f|_b = \dd_\mu f|_w + 2{a,_\mu\over a}b^i{\dd f\over \dd b^i} 
	~,\]
we see, that the  braces in Eq.~(\ref{lstar}) just correspond to 
$\dd_\mu f|_w$. Therefore,
\[a^2X_g f(x,p)=w^\mu\dd_\mu f|_w-\tilde{\Ga}_{\al\beta}^iw^\al w^\beta 
	{\dd f\over \dd w^i} = X_{\tilde{g}}f(x,ap) ~. \]

We have thus shown that the Liouville equation in a perturbed 
Friedmann universe is equivalent to the Liouville equation in perturbed
Minkowski space,
\be (X_{\tilde{g}}f)(x,v)=0 ~, 
	\label{LM} \ee
with $v=v^\mu\tilde{e}_\mu = ap^\mu\tilde{e}_\mu$.\footnote{Note that 
also Friedmann universes with non vanishing spatial curvature, 
$K\neq 0$, are conformally flat and thus 
this procedure can also be applied for $K\neq 0$. Of course, in this
case the conformal factor $a^2$ is no longer just the scale factor but
 depends on position. A coordinate transformation which transforms the
metric of $K\neq 0$ Friedmann universes into a conformally flat form
can be found, e.g., in \cite{CDD}.}

We now want to derive a perturbation equation for Eq.~(\ref{LM}).
If $\bar{e}^\mu$ is a tetrad in Minkowski space, 
$\tilde{e}_\mu = \bar{e}_\mu + {1\over 2}h_\mu^\nu\bar{e}_\nu$ is a 
tetrad w.r.t the 
perturbed geometry $\tilde{g}$. For 
$(x,v^\mu\bar{e}_\mu )\in  \bar{P}_0$, thus,
$(x,v^\mu\tilde{e}_\mu)\in \tilde{P}_0$. Here $\bar{P}_0$ denotes the 
zero mass one particle phase space in Minkowski space and $\tilde{P}_0$ 
is the phase space with respect to $\tilde{g}$, perturbed Minkowski 
space. We define the 
perturbation of the distribution function $F$ by
\be 
   f(x,v^\mu \tilde{e}_\mu) = \bar{f}(x,v^\mu \bar{e}_\mu) + 
	F(x,v^\mu\bar{e}_\mu)
  ~. \ee
Liouville's equation for $f$ then leads to a perturbation equation 
for $F$. We choose the natural tetrad 
\[\tilde{e}_\mu=\dd_\mu -{1\over 2}h_\mu^\nu\dd_\nu\]
with the corresponding basis of 1--forms
\[\tilde{\th}^\mu=dx^\mu +{1\over 2}h^\mu_\nu dx^\nu ~.\]
Inserting this into the first structure equation, 
$d\tilde{\th}^\mu= -\om^\mu_{~~\nu}\wedge dx^\nu$, one finds
\[ \om_{\mu\nu}=-{1\over 2}(h_{\mu\la},_\nu - 
h_{\nu\la},_\mu)\th^\la ~.\]
Using the background Liouville equation, namely that $\bar{f}$ is
only a function of $v=ap$, we obtain the perturbation equation
\[ (\dd_t +\ga^i\dd_i)F = -{v\over 2}[(\dot{h}_{i0}-h_{00},_i)\ga^i
   +(\dot{h}_{ij}-h_{0j},_i)\ga^i\ga^j]{d\bar{f}\over dv}  ~,\]
where we have set $v^i=v\ga^i$, with $v^2=\sum_{i=1}^3(v^i)^2$.
Let us parameterize the perturbations of the metric by
\begin{equation} \left(h_{\mu\nu}\right) = \left(\begin{array}{ll} 
		-2A & B_i \\
                B_i & 2H_L\delta_{ij}
                                +2H_{ij} \end{array}\right),
 \label{scalar} \end{equation}
with $H_i^i=0$. Inserting this above we obtain
\be
(\dd_t +\ga^i\dd_i)F = -[\dot{H}_L +(A,_i +{1\over 2}\dot{B}_i)\ga^i +
	(\dot{H}_{ij}-{1\over 2}B_{i,j})\ga^i\ga^j]v{d\bar{f}\over dv} ~.
 \label {LF} \ee
From Eq.~(\ref{LF}) we see that the perturbation in the distribution 
function in each spectral band is proportional to $v{d\bar{f}\over dv}$.
This shows ones more that gravity is achromatic. We thus
do not loose any information if we integrate this equation over 
photon energies. We define 
\[ m = {\pi\over \rho_ra^4}\int Fv^3dv ~.\]
$4m$ is the fractional perturbation of the brightness $\iota$,
\[ \iota = a^{-4} \int f v^3dv ~. \]
This is obtained using the relation
\be   
{4\pi}\int {d\bar{f}\over dv}v^4dv = -4\int \bar{f}v^3dvd\Om 
	=-4\rho_ra^4    ~. \label{rel}	\ee
Setting $\iota = \bar{\iota}(T(\ga,x))$, one finds that 
$ \iota =(\pi/60) T^4(\bm{\ga},x)$. Hence, $m$  
corresponds to the fractional perturbation in the temperature, 
\be T(\ga,x) = \bar{T}(1+m(\ga,x)) ~.\label{T} \ee
Another derivation of Eq.~(\ref{T}) is given in \cite{d94}.
Since the $v$ dependence of  $F$ is of the form 
$v{d\bar{f}\over dv}$, we have with Eq.~(\ref{rel})
\[ F(x^\mu,\ga^i,v)=-m(x^\mu,\ga^i)v{d\bar{f}\over dv} ~. \] 
This shows that $m$ is indeed the quantity which is measured in a
CMB anisotropy experiment, where the spectral information is used
to verify that the spectrum of perturbations is the derivative of a
blackbody spectrum. Of course, in a real experiment located at a fixed
position in the Universe, the monopole and 
dipole contributions to $m$ cannot be measured. They cannot be 
distinguished from a background component and from a dipole due to our
peculiar motion w.r.t. the CMB radiation.

Multiplying Eq.~(\ref{LF}) with $v^3$ and integrating over $v$,  we obtain
the equation of motion for $m$
\be 
	\dd_tm+\ga^i\dd_im= \dot{H}_L +(A,_i +{1\over 2}\dot{B}_i)\ga^i +
	(\dot{H}_{ij}-{1\over 2}B_i,_j)\ga^i\ga^j ~.
\label{Lm}
\ee

It is well known that the equation of motion for photons only couples to
the Weyl part of the curvature (null geodesics are conformally invariant).
The r.h.s. of Eq.~(\ref{Lm}) is given by first derivatives of the metric only
which could at most represent integrals of the Weyl tensor. To obtain 
a local, non integral equation, we thus rewrite Eq.~(\ref{Lm}) in terms of
$\lap m$. It turns out, that the most suitable variable is however not $\lap m$
but $\chi$, which is given by
\[ \chi = \lap m - (\lap H_L-{1\over 2}H,_{ij}^{ij}) - 
	{1\over 2}(\lap B_i-3\dd^j\si_{ij})\ga^i  ~, \]
\[ \mbox{where }~~ \si_{ij}= -{1\over 2}(B_i,_j+B_j,_i) +
	{1\over 3}\de_{ij}B_l^{,l} +\dot{H}_{ij}.  \]
Note that $\chi$ and $\lap m$ only differ by the monopole contribution,
$\lap H_L-(1/2)H^{ij},_{ij}$ and the dipole contribution, 
$(1/2)(\lap B_i -3\dd^j\si_{ij})\ga^i$. The higher multipoles of
$\chi$ and $\lap m$ agree.
An observer at fixed position and time cannot distinguish a monopole 
contribution from an isotropic background
 and a dipole contribution from a peculiar motion.
Only the higher multipoles, $l\ge 2$ contain information about
temperature anisotropies. For a fixed observer therefore, we can 
identify $\lap^{-1}\chi$ with $\de T/T$.

In terms of  metric perturbations, the electric and magnetic part of the 
Weyl tensor are given by ({\sl see}, e.g. \cite{Ma,d94})
\bea
 E_{ij} &=&  {1\over 2}[\lap_{ij}(A-H_L) -\dot{\si}_{ij}
		-\lap H_{ij}-{2\over 3}H_{lm}^{,lm}\de_{ij}
	+ H_{il}^{,l},_j + H_{jl}^{,l},_i] \label{E} \\
 B_{ij} &=& -{1\over 2}(\ep_{ilm}\si_{jm},_l + \ep_{jlm}\si_{im},_l ) ~,
  \label{B}  \eea
\[ \mbox { with }~~ \lap_{ij} =\dd_i\dd_j -(1/3)\de_{ij}\lap ~.\]

Explicitly working out $(\dd_t+\ga^i\dd_i)\chi$ using Eq.~(\ref{Lm}),
 yields after some algebra the equation of motion for $\chi$:
\be
 (\dd_t +\ga^i\dd_i)\chi = 3\ga^i\dd^jE_{ij} +\ga^k\ga^j\ep_{kli}
	\dd_lB_{ij} \equiv S_T(t,\bm{x},\bm{\ga})~ ,
  \label{Lchi}  \ee
where $\ep_{kli}$ is the totally antisymmetric tensor in three 
dimensions with $\ep_{123}=1$.
The spatial indices in this equation are raised and lowered with 
$\de_{ij}$ and thus  index positions are irrelevant. Double indices are
summed over irrespective of their positions.

In eqn. Eq.~(\ref{Lchi}) the contribution from the electric part of the
Weyl tensor does not contain tensor perturbations.  On the other hand, 
scalar perturbations do not induce a magnetic gravitational field. The
second contribution to the source term in Eq.~(\ref{Lchi}) thus represents a 
combination of vector and tensor perturbations. If vector perturbations
are negligible, the two terms on the r.h.s of Eq.~(\ref{Lchi}) 
yield a split into scalar and tensor  perturbations which is local.

Since the Weyl tensor of Friedmann Lema\^{\i}tre universes vanishes, the
r.h.s. of Eq.~(\ref{Lchi}) is manifestly gauge invariant (this is the so 
called Stewart--Walker lemma \cite{SW}). Hence also the variable
$\chi$ is gauge invariant. 
Another proof of the gauge 
invariance of $\chi$, discussing the behavior of $F$ under infinitesimal
coordinate transformations is presented in \cite{d94}.

The general solution to Eq.~(\ref{Lchi}) is given by

\be 
	\chi(t,\bm{x},\bm{\ga}) =
	\int_{t_i}^t S_T(t',\bm{x}+(t'-t)\bm{\ga}, \bm{\ga})dt'
	~ + ~ \chi(t_i,\bm{x}+(t_i-t)\bm{\ga}, \bm{\ga}) ~,
\label{chi} \ee
where $S_T$ is the source term on the r.h.s. of Eq.~(\ref{Lchi}).
Let us compare this result with the more familiar one, where one
calculates $\de T/T$ by integrating photon geodesics (which is of
course equivalent to solving the Liouville equation). For simplicity,
we specialize to the case of pure scalar perturbations (the expressions
for vector and tensor perturbations given in \cite{d94} can be compared
with Eq.~(\ref{chi}) in the same manner.) For scalar perturbations, 
integration of photon geodesics yields \cite{d94}
\be 
	{\de T\over T}(t_f,\bm{x}_f,\bm{n}) = -[{1\over 4}D_g^{(r)} +
	 V_i\cd n^i + (\Psi-\Phi)]^f_i +
		\int_i^f(\dot{\Psi}-\dot{\Phi})d\la ~. \label{geo}
\ee
Here $\Psi$ and $\Phi$ denote the Bardeen potentials as defined, e.g.,
 in \cite{KS,d94}. 
On super horizon scales (which are the important scales for the
Sachs--Wolfe) contribution $V_i\cd n^i$ can be neglected. Furthermore,
the contributions in the square bracket of Eq.~(\ref{geo})
from the final time $t=t_f$, only lead to
uninteresting monopole and dipole terms. We now use that 
the electric contribution to the Weyl
tensor for purely scalar perturbations is given by \cite{d94})  
\[ E_{ij} = {1\over 2}(\dd_i\dd_j -{1\over 3}\lap)(\Psi-\Phi)
	\equiv {1\over 2}\lap_{ij}(\Psi-\Phi) ~.\]
Therefore $\dd_i(\Psi-\Phi)=3\dd^jE_{ij}$. Using furthermore
\[ -(\Psi-\Phi)\left|^f_i  \right. = -\int_i^f[\dot{\Psi}-\dot{\Phi}+
	(\Psi-\Phi),_in^i]d\la ~, \]
Eq.~(\ref{geo}) leads to
\be 
	{\de T\over T}(t,\bm{x},\bm{n}) = 
	{1\over 4}D_g^{(r)}(t_i,\bm{x}_i) -
	3\int_i^f\lap^{-1}\dd^jE_{ij}n^i dt ~ \label{geo2}
\ee
If we take into account that the direction $\bm{n}$ in Eq.~(\ref{geo}),
the direction of an {\sl incoming } photon corresponds to $-\bm{\ga}$
in Eq.~(\ref{chi}), we find that Eq.~(\ref{geo}) coincides with
Eq.~(\ref{chi}) for scalar perturbations, and that
\be \chi(t_i,\bm{x}_i,\bm{\ga})= {1\over 4}\lap D_g^{(r)}(t_i,\bm{x}_i) 
     ={1\over 4}\lap D_g^{(r)}(t_i,\bm{x} -\bm{\ga}(t-t_i)) ~.
\ee
We now want to investigate this initial value and decompose 
Eq.~(\ref{geo2}) into terms due to CDM and terms coming from the 
source, the scalar field. We assume that dark matter and radiation 
perturbations are  adiabatic  on {\em superhorizon scales}, 
\[D_g^{(r)}=(4/3)D_g^{(c)}~.\]
 Since
radiation and CDM probably have been a single fluid at early times
(e.g. at the time of the phase transition), this assumption is 
reasonable.
It is however, inconsistent to set $D_g^{(r)} =4/3D_g^{(c)}$ on 
subhorizon scales. Due to the different equations of state  for the 
two components, adiabaticity cannot be maintained on sub-horizon 
scales \cite{KS}. 
We can then derive from the equations given in \cite{d94}
\[{1\over 4}D_g^{(r)} ={5\over 3}\Phi_C +
	{2\over 3}\dot{\Phi}_C/(\dot{a}/a) +\Phi^S  ~.\]
Here the Bardeen potentials are split into parts due to 
cold dark matter ($_C$) and the scalar field ($_S$) respectively.
For cold dark matter $\Psi_C=-\Phi_C$. Using this, we can bring 
Eq.~(\ref{chi}) into the form
\bea {\de T\over T}(t_f,\bm{x}_f,\bm{n})  &=&
	{1\over 3}\Psi_C(t_i,\bm{x}_i) -
	{2\over 3}\dot{\Psi}_C/(\dot{a}/a)(t_i,\bm{x}_i)
	+2\int_i^f\dot{\Psi}_C dt  \nonumber \\   
  &&	+\Phi^S(t_i,\bm{x}_i) -
	3\int_i^f\lap^{-1}S_{TS}(t,\bm{x}_f-(t_f-t)\bm{n},\bm{n}) dt 
	~, \label{genu}
\eea
where $S_{TS}$ denote the portion of the source term due to the
scalar field only:
\be 
	S_{TS}= -3n^i\dd^jE^{(S)}_{ij} +n^kn^j\ep_{klj}
	\dd_lB^{(S)}_{ij} ~.  \label{STS}
\ee
Eq. (\ref{genu}) is much better suited
for numerical investigation than the general expression Eq.~(\ref{chi}).
This can be demonstrated by considering the case of pure CDM without
source term: In this case $\Phi_C=-\Psi_C=$ constant and from 
Eq.~(\ref{genu}) we easily recover the well--kown result 
\[ {\de T\over T}(t,\bm{x},\bm{n}) = 
	{1\over 3}\Psi_C(t_i,\bm{x}-\bm{n}(t-t_i)) ~,\]
whereas Eq.~(\ref{chi}) in this case  leads to
\[ {\de T\over T}(t,\bm{x},\bm{n}) = {\de T\over T}(t_i,\bm{x}_i,\bm{n})
	+ 2\Psi_C(t_i,\bm{x}_i) ~.\]
In other words, the unknown initial condition in Eq.~(\ref{chi}) 
cancels $5/6$ of the naive result for the case of adiabatic CDM
fluctuations. Even though due to the existence of $\dot{\Psi}_C$ terms.
the cancelation is slightly less substantial in our case, the 
assumption of adiabaticity on superhorizon scales is a crucial 
ingredient of the model.

The electric and magnetic parts of the Weyl tensor are determined by
the perturbations in the energy momentum tensor via Einstein's
equations. 
We assume that the source for the geometric perturbations is given by
the scalar field and  dark matter. The contributions from radiation
may be neglected. Furthermore, vector perturbations of  dark matter
(which decay quickly) are neglected. 
The  divergence of $E_{ij}$ is then determined  by
({\sl see} Appendix~A)

\be 	3\dd^jE_{ij} = 8\pi G\rho_{C}a^2D_i + 
	8\pi G(\dd_i\de T_{00} +3({\dot{a}\over a})\de T_{0i}
	- (3/2)\dd^j\tau_{ij})
	 \label{dE} ~,
\ee
where the first term on the r.h.s. is the dark matter source term,
$\rho_{C}$ denoting the dark matter energy density. The second
contribution is due to the scalar field: The energy momentum tensor of
the scalar field 
\[ T_{\mu\nu}^{S} = \phi,_\mu\phi,_\nu -
	{1\over 2}g_{\mu\nu}\phi^{,\la}\phi,_\la \]
yields
\[\tau_{ij} \equiv T_{ij} -(a^2/3)\de_{ij}T^l_l =
		\tau_{ij}^{S} 
  = \phi,_i\phi,_j -(1/3)\de_{ij}(\nabla\phi)^2 ~,\]
\[ \de T_{0j}=  T_{0j}^{S} = \dot{\phi}\phi,_j ~~, \]
\[ \de T_{00}=  T_{00}^{S} = {1\over 2}((\dot{\phi})^2
	 + (\nabla\phi)^2) ~,\]
and $D_j$ is a gauge invariant perturbation variable  for the density
gradient. For scalar perturbations $ D_j = \dd_jD$.
The evolution equation for the dark matter density perturbation  
is given by ({\sl see} \cite{d90}) 
\be 
   \ddot{D} + ({\dot{a}\over a})\dot{D} - 4\pi Ga^2\rho_{C}D
	= 8\pi G \dot{\phi}^2
\label{C} ~. \ee
During the radiation dominated era $8\pi G\rho_RD_R$ in principle 
has to be included in Eq.~(\ref{C}). But since radiation 
perturbations quickly decay 
 on sub-horizon scales, and since dark matter fluctuations cannot grow
in a radiation dominated universe \cite{Me}, their influence is not 
relevant. 

The equation of motion for $B_{ij}$ is  more involved. A somewhat 
cumbersome derivation ({\sl see Appendix~A}) yields

\be
	a^{-1}(aB_{ij})^{\cd\cd} -\lap B_{ij}
	= 8\pi G{\cal S}^{(B)}_{ij} ~, \label{Bij}
\ee
\[ \mbox{with }~~~ {\cal S}^{(B)}_{ij}= -\ep_{lm(i}\de T_{0l},_{j)m}
	+\ep_{lm(i}\dot{\tau}_{j)l},_m  ~.\]
Here $(i...j)$ denotes symmetrization in the indices $i$ and $j$.

 To the source term ${\cal S}^{(B)}$
only vector and tensor perturbations contribute. It is thus entirely
determined by the energy momentum tensor of the scalar field.

\section{The System of equations for Global Scalar Field Induced 
Fluctuations}
In this section we collect all the equations which determine the
system under consideration. We also repeat equations which have
already been derived in Section~2. Let us begin with the scalar field
equation of motion.

The energy momentum tensor of the scalar field is a small perturbation.
In first order perturbation theory, we can thus solve the equation
of motion of the scalar field in the background, Friedmann--Lema\^{\i}tre
geometry, neglecting geometric perturbations.
The equation of motion for the scalar field $\phi$ is then given by
\be 
	g^{\mu\nu}\nabla_\mu\nabla_\nu\phi + {\dd V\over \dd\phi} = 0 ~,
  \label{phi1}\ee
where $g^{\mu\nu}$ denotes the unperturbed metric and $\nabla_\mu$ is
the covariant derivative with respect to this metric. For our numerical
computations, we consider an $O(4)$ model. In $O(N)$ models  the
 scalar field,
$\phi\in \bm{ R}^N$ and the zero temperature potential is given by
$V_0={\la\over 4}(\phi^2-\eta^2)^2$ for some energy scale $\eta$. At high
temperatures, $T>T_c \sim \eta$, one loop corrections to the effective
potential dominate and the minimum of the effective potential is at
$\phi=0$. Below the critical temperature the minimum is shifted (in the
simplest case) to $<\phi^2> =(1-(T/T_c)^2)\eta^2$ ({\sl see} \cite{Ki,d95}
and references therein). The vacuum manifold, i.e. the space of minima 
of the effective potential, then becomes a 
$(N-1)$--sphere, $\bm{ S}^{(N-1)}$. Since 
\[	\pi_k(\bm{ S}^m) = \left\{ \begin{array}{ll}
	0~~, & k<m\\
       \bm{ Z}~~, k=m, \end{array}\right.  \]
the lowest non--vanishing homotopy group of a $m$--sphere is always 
$\pi_m$. Since probably higher defects are unstable and decay 
into lower ones\footnote{This is an unproven conjecture, motivated, 
e.g., by observations of the density of textures and monopoles in
liquid crystals and by numerical experiments \cite{CDTY,Kinew}}, the 
$m$--sphere is a suitable vacuum manifold to study $\pi_m$ defects.

If the system under consideration is at a temperature $T$ much below the
critical temperature, $T\ll T_c$, it becomes more and more improbable for
the field $\phi$ to leave the vacuum manifold. $\phi$ will leave the 
vacuum manifold only if it would otherwise be forced to gradients of
order $ (\nabla \phi)^2 \sim \la\phi^2\eta^2$, thus only over length scales
of order $l=1/(\sqrt{\la}\eta) \equiv m_\phi^{-1}$ ($l$ is the transversal
extension of the defects). 
For GUT scale phase transitions  $l \sim 10^{-30}$cm
as compared to cosmic distances of the order of Mpc $\sim 10^{24}$cm.
If we are willing to loose the information of
the precise field configuration over these tiny regions, 
it seems well justified to fix $\phi$ to the vacuum manifold $\cal N$.
Instead of discussing the field equation Eq.~(\ref{phi1}),
we require $ \phi/\eta \in \bm{ S}^{(N-1)}$. The remaining field
equation, $\Box\phi=0$, then  demands that
\[ \phi/\eta\equiv \beta ~:~ {\cal M} \ra \bm{ S}^{(N-1)}  \]
is a harmonic map from spacetime $\cal M$ into {\bf S}$^{(N-1)}$. 

The topological defects we are interested in are  singularities of these
maps. When the gradients of $\phi$ become very large, like, e.g., towards
the center of a global monopole, the field leaves the vacuum manifold
and assumes   non vanishing potential energy. If 
$\beta\in \bm{ S}^{(N-1)}$ is enforced, a singularity develops by 
topological reasons.

In the physics literature harmonic maps are known as $\si$--models.
The action of a $\si$--model is given by
\be
 S_\si = \int_{\cal M}g^{\mu\nu}\dd_\mu\beta^A\dd_\nu\beta^B\ga_{AB}(\beta)
		\sqrt{|g|}d^4x  ~, \label{Ssi}
\ee
where $\ga_{AB}$ denotes the metric on $\bm{ S}^{N-1}$ and 
$g_{\mu\nu}$ is the metric of spacetime.
We now fix $\beta$ to lay in the vacuum manifold, $\bm{ S}^{N-1}$
 by introducing a Lagrange multiplier. We then obtain the 
following equation of motion for $\beta$:
\be
	\Box\b - (\b\cd\Box\b)\b =0 ~,  \label{si2}
\ee
which shows that the $\si$--model is scale free.
There are thus two possible evolution equations for the scalar field 
at low temperature. We call Eq.~(\ref{phi1}) the 'potential model' evolution
equation and Eq.~(\ref{si2}) the $\si$--model approach.

The energy momentum tensor of the scalar field perturbs spacetime
geometry and induces perturbations in the dark matter energy density
according to Eq.~(\ref{C}) 
\be 
   \ddot{D} + ({\dot{a}\over a})\dot{D} - 4\pi Ga^2\rho_{C}D
	= 8\pi G \dot{\phi}^2
\label{D..} ~, \ee
where $D$ is a gauge invariant variable for the dark matter 
perturbations \cite{d90}. On subhorizon scales $D \sim \de\rho/\rho$.
In comoving coordinates, the total perturbed energy momentum tensor is 
given by
\[ \de T_\mu^\nu = \phi,_\mu\cd\phi^{,\nu}-{1\over 2}\de_\mu^\nu
	\phi,_\la\cd\phi^{,\la} +\rho_{C}D\de_\mu^0\de_0^\nu ~.\]
As already mentioned in section~2, the perturbed Einstein equations to
this energy momentum tensor
yield an algebraic equation for the divergence of the electric part
of the Weyl tensor and an evolution equation for the magnetic part
of the Weyl tensor ({\sl see Appendix~A}):
\be 	\dd^jE_{ij} = -{8\pi\over 3} G\rho_{C}a^2D_i - 
	8\pi G({1\over 3}\dd_i\de T_{00} +({\dot{a}\over a})\de T_{0i}
	+ {1\over 2}\dd^j\tau_{ij})
	 \label{djE} ~,\mbox{ and}
\ee
\be
	{1\over a}(aB)_{ij})^{\cd\cd} +  -\lap B_{ij}
	= 8\pi G{\cal S}^{(B)}_{ij} ~, \label{B..}
\ee
\[ \mbox{with }~~~ {\cal S}^{(B)}_{ij}= 
       \ep_{lm(i}[T^{S}_{0l},_{j)m} +
	\dot{\tau}_{j)l},_m]~, ~~~\mbox{ and }~~~~
      \tau_{ij}=\phi,_i\phi,_j-{1\over 3}\de_{ij}(\nabla\phi)^2 ~.\]
The source term for the perturbation of Liouville's equation is given
by Eq.~(\ref{STS}):
\be -3n^i\dd^jE^{(S)}_{ij} + n^kn^j\ep_{klj}
	\dd_lB^{(S)}_{ij} \equiv S_{ST}(t,\bm{x},\bm{n})~ . 
\label{sou} \ee
The CMB fluctuations are then determined according to 
\bea 
	{\de T\over T}(t,\bm{x},\bm{n}) =
	\int_{t_i}^t S_{ST}(t',\bm{x}+(t'-t)\bm{n}, \bm{n})dt'
	~ + ~ \Phi_S(t_i,\bm{x}+(t_i-t)\bm{n}, \bm{n}) ~,
	 \nonumber \\ 
	+	{1\over 3}\Psi_C(t_i,\bm{x}_i) -
	{2\over 3}\dot{\Psi}_C/(\dot{a}/a)(t_i,\bm{x}_i)
	+2\int_i^f\dot{\Psi}_C dt 
\label{chi2} \eea
Eqs.~(\ref{phi1}) and (\ref{D..})  to (\ref{chi2}) form a
closed, hyperbolic  system of partial differential equations. Actually
all except the scalar field equation Eq.~(\ref{phi1}) are linear perturbation
equations with source term and can thus be solved, e.g., by the
Wronskian method, i.e., by some integrals over the source term.
The corresponding solution for $\de T/T$ is given above in
Eq.~(\ref{chi2}),
the general solution of the dark matter equation is given below in
Eqns.~(\ref{D2}), (\ref{Dgen}) and (\ref{wrons}).

Let us briefly describe the general solution for $B_{ij}$: We switch to
Fourier space, because there the $\lap$ is a simple 
multiplication by $-k^2$ and Eq.~(\ref{B..}) becomes an ordinary
differential equation with scalar homogeneous solutions
\be
   b^{\pm} = {1\over a}\exp(\pm ikt) ~.
\ee
The general solution to the inhomogeneous equation is given by
\be 
 B_{ij}=(b^{+}C^{+}_{ij} +b^{-}C^{-}_{ij}) + B^{(hom)}_{ij} 
~,  \label{BW}\ee
where $B^{hom}$ denotes an arbitrary homogeneous solution and
$C^{+}$, $C^{-}$ are given by
\bea
 C^{+}_{ij} &=& -8\pi G\int{\tilde{\cal S}^{(B)}_{ij}b^- \over W}dt\\
C^{-}_{ij} &=& 8\pi G\int{\tilde{\cal S}^{(B)}_{ij}b^+ \over W}dt~.
\eea
Here $W$ denotes the Wronskian determinant of the solutions which 
amounts to
\be
 W=b^+\dot{b}^--b^-\dot{b}^+ = {2ik\over a^2} ~.
\ee

\section{Initial Conditions and Numerical Methods}

\subsection{The scalar field:}
As already shown in the previous section, the equation of motion of the 
scalar field  is given 
by
\be 
	g^{\mu\nu}\nabla_\mu\nabla_\nu\phi + {\dd V\over \dd\phi} = 0 ~,
  \label{phi}\ee
where $g^{\mu\nu}$ is the background (unperturbed metric). 
With $\beta = \phi/\eta$ and $m=\sqrt{\la}\eta$, Eq.~(\ref{phi}) 
yields for $O(N)$ models in a Friedmann universe
\be
  \dd_t^2\beta + 2(\dot{a}/a)\dd_t\beta -\nabla^2\beta = 
	{1\over 2}a^2m^2(\beta^2-1)\beta ~. \label{bb}
\ee
This equation as it stands can not be treated numerically in the regime
which is interesting for large scale structure formation. The two scales
in the problem are the horizon scale $t\sim (\dot{a}/a)^{-1}$ and the
inverse symmetry
breaking scale, the comoving scale $(am)^{-1}$. At recombination, e.g.,
these scales differ by a factor of about $10^{53}$ and can thus not 
both be resolved in one computer code.

There are two approximations to treat the scalar field numerically. 
As we shall see, they are complementary and thus the
fact that both approximations agree with each other within about 10\% 
is reassuring.
The first possibility is to replace $(am)^{-1}$ by $w$, the smallest 
scale which can be resolved in a given simulation, typically twice the 
grid spacing, $w \sim 2\De x$. The time dependence of $(am)^{-1}$ which
 results in a steepening of the potential is mimicked by an additional 
damping term: $ 2(\dot{a}/a) \ra \al \dot{a}/a$, with 
$\al \sim 3$ \cite{PSR}.
Numerical tests have shown, that this procedure, which usually is 
implemented by a modified staggered leap frog scheme \cite{NR}, is not
very sensitive on the values of $\al$ and $w$ chosen. With this method we
have replaced the growing comoving mass $am$ by the largest mass which
our code can resolve. For a $(256)^3$ grid which  simulates
the evolution of the scalar field until today, we obtain
$256\De x \sim t_0 \sim 4\times10^{17}$sec$/a_0$, so that 
$w\sim 4\times 10^{15}$sec$/a_0$,
i.e., $am\sim \eta a_{rec}\sim 10^{17}GeV$ is replaced by about 
$w^{-1}= a_010^{-39 }GeV \sim 10^{-35}GeV$, where we set 
$a_{eq}=1$.

 We are confident that this modified equation  mimics the behavior of 
the field, since the actual mass of the scalar field is irrelevant as 
long as it is much larger 
than the typical kinetic and gradient energies associated with the field
which are of the order the inverse horizon scale. Therefore, as soon as
the horizon scale is substantially larger than $\De x$, the code should
mimic the true field evolution on scales larger then $w$. But, to our
knowledge, there exists no rigorous mathematical approximation scheme 
leading to the above treatment of the 
scalar field which would then also yield the optimal choice for $\al$.

Alternatively, we can treat the scalar field  in the $\si$--model 
approximation given in the previous section. This approach is opposite 
to the one outlined above in 
which the scalar field mass is much too small, since the $\si$--model
corresponds to setting the scalar field mass infinity. 

The $\si$--model  equation of motion cannot be treated numerically 
with a leap frog scheme, since it involves non--linear time
derivatives. In this case, a second order accurate integration scheme
has been developed by varying the discretized action with respect to the
field \cite{PST}.

The two different approaches have been extensively tested by us and
other workers in the field, and  good
agreement has been found on scales larger than about 3 -- 4 grid sizes
\cite{BCL1,B}. We have compared our potential code with the exact 
spherically symmetric scaling solution  \cite{TS} and with our old 
spherically symmetric $\sigma$--model code \cite{DHZ}. Outside the 
unwinding events which extend over approximately 3 grid sizes, the 
different approaches agree within about 5\%.
This is very encouraging, especially since the two 
treatments are complementary: In the $\si$--model, we let the scalar 
field mass $m$ go to infinity. In the potential approach, we replace 
$m$ by 
$\sim 1/\De x \sim 200/t_{0} \sim 200a_0/10^{10}y \sim 10^{-35}$GeV.

The integration of the scalar field equation is numerically the hardest 
part of the problem, since it involves
the solution of a system of nonlinear partial differential equations.
A good test of our numerical calculations, next to checking the 
scaling behavior of $\rho_S$, is energy momentum conservation of the
scalar field, $T^{(S)\mu\nu}_{;\nu}=0$. 
Energy momentum conservation in the potential model, about 15\%, is 
slightly worse than in the $\si$--model, where it is about 5\% 
({\sl see Fig.~2}).
Therefore, the final results presented here are all obtained with
the $\si$--model approach. Our checks lead us to the
conclusion, that we can calculate the scalar field energy momentum
tensor, which then is the source of dark matter and CMB fluctuations 
to an accuracy of  about 10\%.  
The problem of choosing the correct initial condition may induce
another (systematical) error in our calculations which we hope
to remain below 20\%. Other sources of error are negligible.

\subsection{Dark matter}
Once the scalar field $\beta(\bm{x},t)$ is known, the dark matter 
perturbations can easily be calculated by either
using the Wronskian method (see below) or some standard ordinary 
differential equation solver. We have performed both methods and they 
agree very well. For later use, we briefly describe the Wronskian 
method.
We normalize the scale factor by
\[ a= {t\over \tau}(1+{1\over 4}t/\tau) ~~~, \mbox{ with} \]
\[ \tau=1/\sqrt{(4\pi G/3)\rho_{eq}}= {t_{eq}\over 2(\sqrt{2}-1)}~.\]
Here $t_{eq}$ denotes the time of equal matter and radiation density,
$ \rho_{rad}(t_{eq})=\rho_{C}(t_{eq})=(1/2)\rho(t_{eq})$. We have 
normalized $a$ such that $a_{eq}=a(t_{eq})=1$.
Transformed to the variable $a$, the dark matter equation Eq.~(\ref{C}) then
yields
\be
{d^2 D\over da^2} + {2+3a\over 2a(1+a)}{d D\over d a} -{3\over 2a(1+a)}D
	= 2\ep \dot{\beta}^2({ da\over dt})^2 = (1+a)S/\tau^2 ~,
	\label{da}\ee
\[ S=2\ep\dot{\beta}^2 ~ \mbox{ and } \ep=4\pi G\eta^2 ~.\]
The  homogeneous solutions to this linear differential equation are
well known \cite{GP}:
\bea D_1 &=& 1+{3\over 2}a   ~,  \label{D1}\\ 
     D_2 &=& (1+{3\over 2}a)[\ln\left({\sqrt{a+1}+1\over\sqrt{a+1}-1}
	\right) -3\sqrt{a+1}]~.	\label{D2}
\eea
The general solution to Eq.~(\ref{da}) is given by
\be 
	D(t)=c_1(t)D_1(t) + c_2(t)D_2(t)  \label{Dgen}  
\ee
with
\be
	c_1 = -\int(SD_2/W)dt \;\;\mbox{ , }\;\;
    	c_2 =  \int(SD_1/W)dt  \;\; .
	\label{wrons} \ee
 \[	W =D_1\dot{D_2} - \dot{D_1}D_2 = 
	{\dot{a}(2a-1)\over a\sqrt{a+1}} = {2a-1\over a\tau}   \] 
 is the Wronskian determinant of 
the homogeneous solutions. The integrals Eq.~(\ref{wrons}) have to be
performed numerically with $S=\ep\dot{\beta}^2$. When discussing the
initial conditions for $D$ in subsection~4.4, we shall present an
analytic approximation for the source term $S$.

\subsection{The CMB anisotropies}
The CMB anisotropies are  given by 
\[ {\de T\over T} = \lap^{-1}\chi \]
up to monopole and dipole contributions which we disregard.
 Here $\chi$ is a solution Eq.~(\ref{chi}) of
 Eq.~(\ref{Lchi}). The source term $S_T$ is determined via Eq.~(\ref{dE})
and Eq.~(\ref{Bij}). However, using this straightforward approach results
in a big waste of computer memory (which we cannot afford): We would
be satisfied to calculate $\de T/T$ for about 30 observers in each 
simulations, which means we need $\lap^{-1}\chi$ only at 30
positions {\bf x}. But since we have to perform an inverse Laplacian
which is done by fast Fourier transforms, we have to calculate $\chi$
on the whole grid, which consists of $192^3 \sim 8\cd 10^6$ positions. In
Addition, to calculate the spherical harmonic amplitudes of $\de T/T$
up to about $l\sim 40$ (angular resolution of about $4^o$), we need 
typically $5\times10^4$ directions $\bm{n}$. The $\chi$ variable alone
(in double precision) would thus require 16 G-bytes of memory, an 
amount which is not available on most present day computers. The way
out is  to take the inverse Laplacian already in the equation of
motion Eq.~(\ref{Lchi}). This results in
\be
 (\dd_t +\ga^i\dd_i){\de T\over T} = -3\ga^i\lap^{-1}(\dd^jE_{ij}) 
- \ga^k\ga^j\ep_{kli}\lap^{-1}(\dd_lB_{ij}) \equiv 
	\lap^{-1}S_T(t,\bm{x},\bm{\ga})~ .
  \label{LT}   \ee
Here the inverse Laplacian has to be performed for a vector field and a 
symmetric traceless tensor field, a total of 8 scalar variables which 
only depend on {\bf x} and not on $\bm{\ga}$. For a $192^3$ grid
this requires nearly 1 Gbyte of memory, no problem for presently
available machines. Eq.~(\ref{LT}) has the general solution ({\sl see}
Eq.~(\ref{genu}))
\bea 
{\de T\over T}(t_0,\bm{x}_0,\bm{\ga}) &=& 
	+\Phi^S(t_i,\bm{x}_i) -
	3\int_i^f\lap^{-1}S_{TS}(t,\bm{x}_0-(t_f-t)\bm{n},\bm{n}) dt 
	 \nonumber \\   &&
	+	{1\over 3}\Psi_C(t_i,\bm{x}_i) -
	{2\over 3}\dot{\Psi}_C/(\dot{a}/a)(t_i,\bm{x}_i)
	+2\int_i^f\dot{\Psi}_C dt 
 ~,\label{dT} \eea
with $S_{TS}$ given in Eq.~(\ref{STS}).
The first term of Eq.~(\ref{dT}) determines the initial condition of
the CMB anisotropies due to the source term. In the numerical simulation
we just set it $ -3n^l\lap^{-1}(\dd^jE^{(S)}_{lj})(t_i,\bm{x}_i)t_i$. 
This assumes that 
the source term is approximately constant until $t_i$ and that 
magnetic contributions can be neglected. The resulting amplitude is not
very sensitive to this assumption, but changing it can somewhat change
in the spectral index.
We have solved Eq.~(\ref{dT}) numerically by just summing
up the contributions from each time step for 27 observer positions
$\bm{ x}_0$.  The value of the source term at  position 
$\bm{x}_0+(t-t_0)$ is determined by linear interpolation. The
quantity $\dd^iE^S_{ij}$ is determined by Eq.~(\ref{dE}) and its inverse 
Laplacian is calculated by fast
Fourier transforms. To obtain $\lap^{-1}B^S_{ij}$ from equation 
Eq.~(\ref{Bij}), we directly calculate
$\lap^{-1}{\cal S}_B$ in $k$--space, then solve the ordinary, linear 
differential equation for $\lap^{-1}B_{ij}$ in $k$--space 
by the Wronskian method. Since, as we shall argue later, all components
$T^{(S)}_{\mu\nu}$ in average scale like $A/\sqrt{t}$ on super horizon 
scales, $S^(B) \sim At^{-3/2}$ and therefore $C^{\pm} \propto t^{3/2}$
on super horizon scales. Therefore, we can neglect the contribution
to $C^\pm$ from the lower boundary in the integral. Furthermore, since 
the homogeneous solution $B_{ij}^{(hom)}$ is decaying, we drop it
entirely. This procedure corresponds to setting $B(t_{i})=0$ and 
calculating $B(t)$ according to Eq~(\ref{BW}).

\subsection{Initial conditions}
Initially, the field $\phi$ itself and/or the velocities $\dot{\phi}$ are
 laid down randomly on the grid points. The initial time, $t_{in}$ is
chosen to be the grid size, $t_{in} = \De x$, so that the field at
different grid points should not be correlated. The configuration is
then evolved in time with one of the approximation schemes discussed 
above.

Because our initial conditions for dark matter and photons very 
sensitively depend on the scaling behavior of the dark matter,
we can only start the dark matter or photon simulations when scaling
is fully reached, $t_{in}=8\De x$. Starting our simulations, e.g., at 
$t=4\De x$ changes the results by about a factor of 2.
Further doubling of the initial time, changes our results by less 
than 20\%, we thus believe that at $t=8\De x$ scaling is sufficiently
accurately. Unfortunately, this late initial time reduces our 
dynamical range to about $192/8=24$ for a $192^3$ grid, which is seen 
clearly in our results for the CMB anisotropies discussed below.

It is very important to choose the correct initial conditions for the
dark matter and, especially,  photon perturbations. Changing them
can change the CMB fluctuation amplitudes by nearly a factor two.  
Since these
fluctuations are used to normalize the model, i.e. to determine $\ep$,
this reflects in corresponding changes in $\ep$. We want to do better
than a factor two by choosing physically plausible initial conditions.
The cleanest way would by to simulate the evolution of perturbations
through the phase transition, assuming that before the phase transition,
the universe was an unperturbed Friedmann universe with $\phi\equiv 0$.
On the other hand, since we want to calculate the  perturbation
spectrum on scales of up to 1000Mpc with a $(256)^3$ grid, 
we cannot start  our dark matter and CMB simulation earlier than at a 
time when the horizon
distance is approximately $8\De x \sim 30$Mpc. At the beginning of the 
scalar field simulation our grid scale $\De x\sim 4$Mpc is of the order
of the horizon scale. We therefore have 
to decide on the amplitudes of super horizon perturbations. One
possibility  is setting all geometrical
perturbations initially to zero. The requirements 
\[	\dd^jE_{ij}(t_{i})=0 ~~\mbox{ and }~~~ S^B_{ij}(t_{i})=0\]
 then yield initial conditions for the dark matter fluctuations $D$ 
and the photon variable $\chi$. But these, let us call them 'strict
isocurvature' initial conditions are not natural since they do 
not propagate in time: Even if we start with $E$ and $B$ vanishing on
super horizon scales, after some time residual fluctuations have leaked
into these scales and one obtains the white noise fluctuations spectrum
on super horizon scales shown in {\sl Fig.~3}.
This does not violate causality, since white noise is uncorrelated
and just results from the residuals of correlated fluctuations on smaller
scales. The correct initial values for $D$ and $\dot{D}$ would of course
be those obtained by solving the equation of motion Eq.~(\ref{C}) from the
symmetry breaking time until the start of the simulation. We found a
method to incorporate this at least approximately: The spectrum of the 
dark matter source term
$8\pi G| \widetilde{\dot{\phi}^2}|^2 $ can be approximated by
\be 
8\pi G |\widetilde{\dot{\phi}^2}| =2\ep \widetilde{\dot{\beta}^2} =
  \ep\sqrt{1\over V}\int d^3x\dot{\beta}^2(x)e^{ikx}  \approx
 {\ep A\over \sqrt{t}(1+a_1kt +a_2(kt)^2)} ~, \label{fit}\ee
with
\[ A= 3.3 ~,~~~ a_1=-0.7/(2\pi)~,~~~ a_2 = 0.7/(2\pi)^2  ~. \]
This numbers have been obtained by a $\chi^2$--minimization scheme. The
approximation is not very good. It yields a $\chi^2 \approx 2000$
for about 1000 data points. Its comparison with the real data
in {\sl Figs.~4 and 5} shows that Eq.~(\ref{fit}) approximates the 
source term
to about 10\% on superhorizon scales, but does not follow the wiggles
present in the data on smaller scales. Since we shall not use the fit
 on subhorizon scales, this is not important for our simulations.
However, in general $\widetilde{\dot{\phi}^2}$ is complex and setting it
equal to its absolute value, we neglect the evolution of phases.
Again, by causality, this will not severely affect scales larger than
the horizon, since on these scales the phases are (approximately) 
frozen. But on subhorizon scales our fis is nor very useful due to
the incoherent evolution of phases.
Assuming this form of the source term, we can solve Eq.~(\ref{C}) 
analytically on super horizon scales, where we  approximate the 
source  term by
\be 2\ep \widetilde{\dot{\beta}^2} =  {\ep A\over \sqrt{t}},
	~~~~~\mbox{ on super horizon scales.} \label{sqrtt}\ee
The homogeneous solutions of Eq.~(\ref{C}) are given by Eq.~(\ref{D1},\ref{D2})
The general inhomogeneous solution, $D=c_1D_1 +c_2D_2$, even with the 
simple source term Eq.~(\ref{sqrtt}), becomes rather complicated. But
in the radiation and matter dominated regimes we find the simple
approximations
\bea
 D  &=& (4/7)t^2S ~~;~~ \dot{D}=(6/7)t^2S  ~~~
	\mbox{ radiation dominated} \label{Dinr} \\
 D  &=& -(4/9)t^2S ~~;~~ \dot{D}=-(2/3)t^2S ~~~\mbox{ matter dominated.} 
	\label{Dinm}\eea

From $D$ we can calculate $\Psi_C$, leading to the dark matter 
contribution to the CMB anisotropies.

As mentioned above, the initial contribution of the scalar field is
approximated by
\[ \Phi^S(t_i,\bm{x}_i) \sim -3t_in^i\lap^{-1}(\dd^jE^{(S)}_{ij}) ~.\]
The result is very sensitive to initial conditions: If we do not 
separate the dark matter and choose some arbitrary, non-adiabatic
initial condition, the resulting $C_\ell$'s increase by nearly a factor
of $10$ and the dark matter induces $80\%$ of the total fluctuation.
However, choosing the adiabatic initial condition discussed in 
Section~3, leading to Eq.~(\ref{dT}), dark matter only contributes about
20\% to the $C_\ell$'s and the main contribution is due to the defects.
The dark matter contribution to the CMB anisotropies is not scale 
invariant, but white noise. It has spectral index $n=0$. This result
was found numerically ({\sl see Fig.~6}) but, as we argue in Appendix~B, it
can also be understood analytically.

our value of $\ep$ obtained with these physical isocurvature and
on super horizon scales adiabatic initial
conditions is in reasonable agreement with the values obtained in
\cite{PST} and \cite{BR}. 

Let us also present a heuristic derivation of the numerical finding 
Eq.~(\ref{sqrtt}) on superhorizon scales: We know that the average value 
  $\langle\dot{\beta}^2\rangle \propto 1/t^2$, the usual scaling 
behavior. The Fourier transform of $\dot{\beta}^2$ determines the
fluctuations on this 'background' on a given comoving 
scale $\la=2\pi/k$. As long as this scale is super horizon, $\la>t$,
a patch of size $\la^3$ consists of $N=(\la/t)^3$ independent horizon
size volumes. The fluctuations on this scale should thus be 
proportional to 
\[\tilde{\dot{\beta}^2}\propto \langle\dot{\beta}^2\rangle /\sqrt{N}  
	\propto 1/\sqrt{t} ~, \]
which is just the behavior which we have found numerically on 
super--horizon scales.

As soon as a given scale becomes sub--horizon, $\la\ll t$,
$\tilde{\dot{\beta}^2}$ starts decaying from this large scale value like
$1/t^2$.

\section{Results}
\subsection{CMB anisotropies}
To analyze the CMB anisotropies, we expand  $\de T/T$
in spherical harmonics
\begin{equation}  {\de T\over T}(t_0,\mbox{\boldmath{$x,\gamma$}}) = 
 \sum_{lm}a_{lm}(\mbox{\boldmath{$x$}})Y_{lm}(\mbox{\boldmath{$\gamma$}})
 ~. \end{equation}
As usual, we  assume that the average over $N_x$ different observer 
positions coincides with the ensemble average and define
\begin{equation} C_\ell = {1\over (2\ell+1)N_x} \sum_{m,x}
	|a_{\ell m}(\mbox{\boldmath{$x$}})|^2 ~, ~~ \ell\ge 2
 \label{cl} ~. \end{equation}
Gaussian fluctuations are characterized by the two point 
correlation function. Since the angular two point correlation
function is given by
\be \langle {\de T\over T}(\bm{ n}){\de T\over T}(\bm{ n}')
     \rangle_{(\bm{ n\cd n}'=\cos\th)} ={1\over 4\pi} \sum_\ell(2\ell+1)
	 C_\ell P_\ell(\cos\th)  ,\ee
Gaussian distributed CMB fluctuations  are fully 
determined by the $C_\ell$'s. 
However, as can be seen from Fig.~7, in our case the distribution of
the CMB fluctuations is not quite Gaussian. It is slightly negatively
skewed. We find an average skewness of $-0.5$ and a kurtosis of
$0.7$.
For a  simulation on a $192^3$ grid with 27 different observer 
positions for each simulation. The  harmonic amplitudes with
are shown in Fig.~8. 
The low order multipoles depend strongly on  the random initial 
conditions (cosmic variance), like in the spherically symmetric 
simulation \cite{DHZ}. 

It is well known, that cold dark matter fluctuations with a power 
spectrum of  spectral index $n$ gravitationally induce CMB anisotropies
with a spectrum given by \cite{Ef}
\be C_\ell = C_2{\Ga(l+(n-1)/2)\Ga((9-n)/2)\over \Ga(l+(5-n)/2)\Ga((n+3)/2)}
	\label{spec} ~.\ee
We have performed a least square fit of our numerical results for 
$\log(C_\ell)$ and the $\log(C_\ell)$ obtained from 
Eq.~(\ref{spec}).\footnote{In the case of
topological defect induced fluctuations, the $C_\ell$ spectrum does not
have exactly this form, since CBM fluctuations are not only induced
by the dark matter but mainly by the scalar field perturbations and the
assumptions made for the derivation of this formula are not valid.}
 If we take into account
all the $C_\ell$'s reliably calculated in our simulations, which 
limits us approximately to $\ell\le 22$ we find a very nice scale 
invariant spectrum,
\be n= 0.9\pm 0.2 \ee
with quadrupole amplitude
\be Q =\sqrt{(5/4\pi)C_2}T_{CMB} = 2.8\pm 0.7 K \cd \ep ~. \label{Q}\ee
 The 1, 2 and 3 sigma contour plot is shown in Fig.~9. The minimal 
$\chi^2$ is 0.56.

It is very interesting, that the dark matter contribution to the CMB 
anisotropies does not yield a scale invariant spectrum, but white noise.
This can be understood analytically: The $\dot{\Psi}_C$ contributions
to $\de T/T$ in Eq.~(\ref{genu}) are not very important and 
\[ (\de T/T)_C(t_0,\bm{k}) \sim
	{1\over 3}\Psi_C(t_i,\bm{k})\exp(i\bm{k\cd nt_0}) =
  {\ep A\over 6\sqrt{t_i}}{\exp(i\bm{k\cd nt_0})\over k^2} \]
on super horizon scales. For the second equal sign we used
$k^2\Psi_C=4\pi GD_C\sim {\ep A\over 2\sqrt{t_i}}$.
By standard arguments ({\sl see, e.g.} \cite{Ef}) one then finds
\be
C_\ell^{(C)} = {\ep^2A^2 \over 18\pi t_i}\int{dk\over k^2}j_\ell^2(kt_0) 
  \propto {\Ga(\ell-0.5)\over \Ga(\ell+2.5)} ~,
\ee
corresponding to Eq.~(\ref{spec}) with $n=0$. This is also what we 
find numerically ({\sl see Fig.~6}). The dark matter contribution
caused the spectral index $n$ of the total CMB anisotropies to drop
slightly below $n=1$.

To reproduce the COBE amplitude $ Q_{COBE}= (20 \pm 5)\mu K$
\cite{Go}, we have to normalize the spectrum by choosing the phase 
transition scale $\eta$ according to
\begin{equation} \epsilon = 4\pi G\eta^2 = (0.8 \pm 0.4)10^{-5} 
	\label{ep}~. \end{equation}
This value is somewhat smaller, but still comparable with the results
obtained in \cite{BR,PST}.

Another method to determine $\epsilon$ is the following: The total
temperature fluctuation amplitude on a given angular scale $\th_c$ is
given by
\be 
	\si^2_T(\th_c) = {1\over 4\pi}
	\sum_\ell C_\ell(2\ell+1)\exp(-\ell^2\th_c^2/2)
	  ~. \ee
In Fig.~10 we show $\si_T$ as a function of $\th_c$. In a recent
analysis of the COBE data\cite{WB} $\si_T^{(COBE)}(7^o) \sim 44\mu K$ and
$\si_T^{(COBE)}(10^o) \sim 40\mu K$ for a spectral index $n\sim 1$,
which leads again to the result given in Eq.~(\ref{ep}).

\subsection{Dark matter fluctuations}
Using fast Fourier transforms we calculate the spectrum
$P(k) = |\delta(k)|^2$ of the dark matter density fluctuations
is shown in Fig.~11. 
The fit shown as solid line in Fig.~11 is given by
 \be 
P(k)h^3/(2\pi)^3= {Ck\over (1 +\al k + (\beta k)^{1.5} +(\ga k)^2)^2}
 ~,	\label{Pk} \ee
with $h=0.5$ and
\bea
	C &=& 215h^{-1}Mpc^4   \label{CC}\\
	\al &=& 10 h^{-2}Mpc  ~~(=0.5\tau) \\
	\beta &=& 1.25h^{-2} Mpc  ~~ \sim \tau/(4\pi)  \\
	\ga  &=& 2.3 h^{-2} Mpc  ~~ \sim \tau/(2\pi) ~,
\eea
where we have used $\tau=19.36h^{-2}$Mpc which is approximately the 
comoving time at equal matter and radiation.

The parameter $C$, which is most important to determine the bias factor
can also be obtained by the following rough analytical argument:
On super horizon scales, $|D|^2 \sim (0.5\ep A)^2t^3$ according to
Eq.~(\ref{Dinm}). As soon as the perturbation enters the horizon at
$t=2\pi/k$, the
source term disappears and $D$ starts growing like $t^2$, leading to
\be
	P(k,t_0) \sim  {(0.5\ep A)^2\over 2\pi}kt_0^4 =
	{(2\pi)^3\over h^3}C_{an}k ~.
\ee
Inserting the numbers $\ep=0.8\times 10^{-5}$, $A=3.3$, 
$t_0^2 =4a_0\tau^2$,
$ a_0\sim 2.5h^2\times 10^4$, we obtain $C_{an}\sim 190$ in excellent
agreement with Eq.~(\ref{CC}). 
Figure~11 can be compared directly with the IRAS observation
\cite{IRAS} and it is compatible with a bias factor of order 1.
A more detailed calculation with Gaussian or square hat window function
yields for $\ep=0.8\times 10^{-5}$
\bea 
\si_{sim}(10MPc)=  1/b_{10} \sim 0.5 - 1  &  \si_{QDOT}(10MPc)=  1&  
	\mbox{for $h=0.5$},  
\eea
yielding $b_{10}\sim 1 - 2 $ for the value of $\epsilon$ found by
comparison with COBE Eq.~(\ref{ep}) A value even somewhat closer to
1 is found for $b_{20}$.
Observations and simulations of nonlinear clustering of dark matter
and baryons \cite{CO} suggest a bias factor $b_{10}\sim 1 - 2$ which is
compatible with our results. It is remarkable, that unlike in the
simulations by Pen et al. \cite{PST}, our bias factor is approximately
constant and  physically acceptable. (To determine our power spectrum,
we have not taken into account 
any smoothing which might change the results by at most 15\%.)
In Fig.~12 we have shown the dark matter pixel distribution from a
$100^3$ simulation. It is interesting, that the skewness of the dark 
matter  distribution is positive, where the $\de T/T$ skewness
is negative.

\section{Conclusions}
Our simulations show that global texture lead to a scale invariant
spectrum of microwave background fluctuations on large scales like
inflationary models of structure formation. This is one of the main
results of this investigation. It is however interesting
that the dark matter contribution to the CMB anisotropies is not scale
invariant, but white noise. Therefore it is important that the initial
condition for dark matter and radiation are adiabatic in which case
the dark matter contribution to the $C_\ell$'s is small and the
flat spectrum caused by the defects is maintained.

Our second main result is the dark matter fluctuation spectrum.
The spectrum is very close to scale invariant and the bias factor
needed for $\ep$ from the CMB anisotropies is around $b\sim$ 1--2. 
This value is certainly acceptable and smaller than the bias factor
obtained in previous investigations \cite{CST}.

The deviation from Gaussian statistics seems to us not very
significant ({\sl see Figs.~7 and 12}) and it is thus important
to develop other means to distinguish topological defects from
inflationary scenarios. A clean and promising candidate for this
distinction are the Doppler peaks which are calculated for the
texture scenario in \cite{DGS}.
 
From our investigations we thus conclude that concerning the large
scale CMB anisotropies and the linear dark matter perturbation spectrum
the texture scenario and probably also other models with global defects
are compatible with present observations.
\vspace{2cm}

\noindent {\large\bf Acknowledgment}
We thank the staff at CSCS, for valuable
support. Especially we want to mention Andrea Bernasconi, Djiordic
Maric and Urs Meier. We acknowledge discussions with Joachim Laukenmann
who  carefully read the manuscript and checked the more involved
algebraic derivations with Maple. We profited from many helpful and
encouraging discussions of this work especially we want to mention
 Marc Hindmarsh, Philipp Jetzer, Yipeng Jing, Mairi Sakellariadou,
Norbert Straumann, Neil Turok, Simon White and others.
We also thank Neil Turok for providing us their $\si$--model code.

\newpage
\appendix
\setcounter{equation}{0}
\renewcommand{\theequation}{A\arabic{equation}}

\section{The equation of motion for the magnetic part of the Weyl 
	tensor} 
The Weyl tensor of a spacetime $({\cal M},g)$ is defined by
\be 
	C^{\mu\nu}_{~\;\;\si\rho}= R^{\mu\nu}_{~\;\;\si\rho} 
	-2g^{[\mu}_{~\;[\si}R^{\nu]}_{~\; \rho]}
	+{1\over 3}Rg^{[\mu}_{~\;[\si}g^{\nu]}_{\;~ \rho]} ~,
	\label{weyl} \ee
where $[\mu ... \nu]$ denotes anti-symmetrization in the indices $\mu$
and $\nu$.
The Weyl curvature has the same symmetries as the Riemann curvature
and it is traceless. In addition the Weyl tensor  is 
invariant under conformal transformations:
\[ C^{\mu}_{\;\;\nu\si\rho}(g)= C^{\mu}_{\;\;\nu\si\rho}(a^2g) \]
(Careful: This equation only holds for the given index position.) 
In four dimensional spacetime, the Bianchi identities together with
 Einstein's equations yield equations of motion for the Weyl curvature.
In four dimensions, the Bianchi identities,
\[ R_{\mu\nu[\si\rho;\la]} = 0 \]
are equivalent to  \cite{CDD}
\be C^{\al\beta\ga\de};_\de = R^{\ga [\al;\beta ]} -{1\over 6}
 g^{\ga [ \al}R^{;\beta ]}  ~. \label{bianchi}\ee
This together with Einstein's equations yields
\be C^{\al\beta\ga\de};_\de = 8\pi G(T^{\ga[\al;\beta]} -{1\over 3}
 g^{\ga[\al}T^{;\beta]})  ~, \label{einstein}\ee
where $T_{\mu\nu}$ is the energy momentum tensor, $T=T^\la_\la$.

Let us now choose some time-like 
unit vector field $u$, $u^2=-1$. We then 
can  decompose any  tensor field into longitudinal and transverse 
components with respect to $u$. We define
\[ h^\mu_{~\nu} \equiv g^\mu_{~\nu} +u^\mu u_\nu ~, \]
 the projection onto the subspace of tangent space normal to $u$.
The decomposition of the Weyl tensor yields its electric and magnetic 
contributions:
\bea E_{\mu\nu} &=& C_{\mu\la\nu\si}u^\la u^\si \\
 B_{\mu\nu} &=& {1\over 2}C_{\mu\la\ga\de}u^\la\
	\eta^{\ga\de}_{\;\;\nu\si} u^\si ~;\eea
where $\eta^{\al\beta\ga\de}$ denotes the totally antisymmetric 4 tensor
with $\eta_{0123}=\sqrt{-g}$.
Due to  symmetry properties and the tracelessness of the Weyl 
curvature, $E$ and $B$ are symmetric and traceless, and they fully 
determine the Weyl curvature. One easily checks that $E_{\mu\nu}$ and
$B_{\mu\nu}$ are also conformally invariant.
We now want to perform the corresponding decomposition for the energy 
momentum
tensor of the scalar field, $\phi$
\[ T^{S}_{\mu\nu} = \phi,_\mu\phi,_\nu -
	{1\over 2}g_{\mu\nu}\phi^{,\la}\phi,_\la ~.\]
We define
\bea
\rho_S &\equiv& T^{(S)}_{\mu\nu}u^\mu u^\nu     \\
p_S &\equiv& {1\over 3}T^{(S)}_{\mu\nu}h^{\mu\nu}     \\
q_\mu &\equiv& -h_\mu^{~\nu}T^{(S)}_{\nu\al}u^\al ~~~ 
	~~~  q_i=-{1\over a}T^{(S)}_{0i} \\
\tau_{\mu\nu} &\equiv& h_\mu^{~\al}h_\nu^{~\beta}T^{(S)}_{\al\beta}-
	h_{\mu\nu}p_S ~. \eea
We then can write
\be T_{\mu\nu}^{(S)}=\rho_Su_\mu u_\nu +p_Sh_{\mu\nu} +q_\mu u_\nu
	+u_\mu q_\nu +\tau_{\mu\nu} ~. \label{Tsplit} \ee
This is the most general decomposition of a symmetric second rank tensor.
It is usually interpreted as the energy momentum tensor of an imperfect
fluid. In the frame of an observer moving with four velocity $u$,
$\rho_S$ is the energy density of the scalar field, $p_S$ is the 
isotropic pressure, $q$ is
the energy flux, $u\cd q=0$, and $\tau$ is the tensor of anisotropic
stresses, $\tau_{\mu\nu}h^{\mu\nu}=\tau_{\mu\nu}u^\mu=0$. 

We now want to focus on
a perturbed Friedmann universe. We therefore  consider
a four velocity field $u$ which deviates only in first order from the
Hubble flow: $u=(1/a)\dd_0 +$ first order. Friedmann universes are conformally
flat, and we require the scalar field to be a small perturbation on a universe
dominated by radiation and cold dark matter (CDM). The energy momentum
tensor of the scalar field and the Weyl tensor are of thus of first order,
and (up to first order) their
decomposition does not depend on the choice of the first order 
contribution to $u$, they are gauge--invariant. But the decomposition
of the dark matter depends on this choice. Cold dark matter is
a pressure-less perfect fluid We can thus choose $u$ to denote the energy
flux of the dark matter, $T^\mu_\nu u^\nu = -\rho_{C} u^\mu$. Then the
energy momentum tensor of the dark matter has the simple decomposition
\be T^{(C)}_{\mu\nu} = \rho_{C}u_\mu u_\nu  \label{DM} ~. \ee
With this choice, the Einstein equations Eq.~(\ref{einstein}) linearized
about an $\Om=1$ Friedmann background with $T^{(S)}_{background} =0$ 
yield the following 'Maxwell equations' for $E$ and $B$ \cite{El}:\\
{\em i) Constraint equations}
\bea  
\dd^iB_{ij} &=& 4\pi G \eta_{j\beta\mu\nu}u^\beta q^{[\mu;\nu]}  
	\label{diB}\\
\dd^iE_{ij} &=& 8\pi G( {1\over 3}a^2\rho_{C}D,_j +{1\over 3}a^2\rho_S,_j 
	-{1\over 2}\dd^i\tau_{ij} -{\dot{a}\over a^2}q_j) ~.
	\label{diE}\eea
{\em ii) Evolution equations}
\bea  
a\dot{B}_{ij} +\dot{a}B_{ij} -a^2h_{(i}^{~~\al}
	\eta_{j)\beta\ga\de}u^\beta E_{\al}^{~~\ga;\de}  
    &=&- 4\pi Ga^2h_{\al(i}\eta_{j)\beta\mu\nu}u^\beta\tau^{\al\mu;\nu}
	  \label{dtB} \\
\dot{E}_{ij} +{\dot{a}\over a}E_{ij} +ah_{(i}^{~~\al}
	\eta_{j)\beta\ga\de}u^\beta B_{\al}^{~~\ga;\de}  
	&=& -4\pi G(aq_{ij}  -{\dot{a}\over a}\tau_{ij} +\dot{\tau}_{ij} 
	+a\rho_{C}u_{ij}) ,
	\label{dtE}\eea
where $(i ...j)$ denotes symmetrization in the indices $i$ and $j$. 
The symmetric traceless tensor fields $q_{\mu\nu}$ and $u_{\mu\nu}$ 
are defined by
\bean
	q_{\mu\nu}&=& q_{(\mu;\nu)}-{1\over 3}h_{\mu\nu}q^\la_{~;\la}\\
  u_{\mu\nu}&=& u_{(\mu;\nu)}-{1\over 3}h_{\mu\nu}u^\la_{~;\la} ~.
\eean
In Eqs.~(\ref{dtB}) and (\ref{dtE})
we have also used that for the dark matter perturbations only scalar
perturbations are relevant, vector perturbations decay quickly. Therefore
$u$ is a gradient field, $u_i =U_{; i}$ for some suitably chosen 
function $U$. Hence the vorticity of the vector field $u$ vanishes,
$u_{[\mu;\nu]}=0$. With
\[ \eta_{0ijk} =a^4\ep_{ijk} ~~,~~~ \rho_S=a^{-2}T^{S}_{00} ~~
	\mbox{ and }~~ q_i = -a^{-1}T_{0i}^{S}  ~,\]
we obtain from Eq.~(\ref{diE})
\be
\dd^iE_{ij} = 8\pi G( {1\over 3}\rho_{C}a^2D,_j +{1\over 3}
	T_{00}^{S},_j
	-{1\over 2}\dd_i\tau_{ij} +{\dot{a}\over a}T_{0j}^{S}) ~.
 \label{dEj}\ee
In Eq.~(\ref{dEj}) and the following
equations summation over double indices is understood, irrespective of 
their position.

To obtain the equation of motion for the magnetic part of the
Weyl curvature we take the time derivative of Eq.~(\ref{dtB}), using
 $u=(1/a)\dd_0 +1.$order and $\eta_{0ijk}=a^4\ep_{ijk}$. This leads to
\be
 (aB_{ij})^{\cd\cd} = -a(\ep_{lm(i}[\dot{E}_{j)l}
	+{\dot{a}\over a}E_{j)l}],_m -4\pi G
	\ep_{lm(i}[\dot{\tau}_{j)l},_m + 
	{\dot{a}\over a}\tau_{j)l},_m])  ~,
\label{ddB} \ee
where we have again used that $u$ is a gradient field and thus terms
like $\ep_{ijk}u_{lj},_k$ vanish.  We now insert Eq.~(\ref{dtE}) into the
first square bracket above and  replace
product expressions of the form $\ep_{ijk}\ep_{ilm}$
and $\ep_{ijk}\ep_{lmn}$ with double and triple Kronecker
deltas. Finally we replace divergences of $B$ with the help of 
Eq.~(\ref{diB}). After some algebra, one obtains
\[ \ep_{lm(i}[\dot{E}_{j)l} +{\dot{a} \over a}E_{j)l}],_m = -\lap B_{ij}
	-4\pi G\ep_{lm(i}[2aq_l,_{mj)} + \dot{\tau}_{j)l},_m -
	{\dot{a}\over a^2}\tau_{j)l},_m] ~.\]
Inserting this into Eq.~(\ref{ddB}) and using 
$aq_l=-T^{S}_{0l}=-\dot{\phi}\phi,_i$,
we finally obtain the equation of motion for $B$:
\be
 a^{-1}(aB)^{\cd\cd}_{ij}   -\lap B_{ij}=	8\pi GS^{(B)}_{ij} ~,
\label{AB}\ee
with
\be
S^{(B)}_{ij} = \ep_{lm(i}[-T^{S}_{0l},_{j)m} +
	\dot{\tau}_{j)l},_m]~, ~~~\mbox{ and }~~~~
      \tau_{ij}=\phi,_i\phi,_j-{1\over 3}\de_{ij}(\nabla\phi)^2 ~.
\ee

 Since dark matter only induces
scalar perturbations and  $B_{ij}$ consists of vector and tensor 
perturbations, it is  independent of the dark matter fluctuations.
Equations Eqs.~(\ref{dEj}) and (\ref{AB}) are used in section~2, where we
need $\dd^iE_{ij}$ and $B_{ij}$ as source terms in the Liouville 
equation.
\newpage

{\Large Figure Captions}
\vspace{1cm}\\
{\bf Fig.~1}\\The scaling behavior for $(\rho+3p)a^2$ found
numerically in $(128)^3$ simulations for different $O(N)$ models. The
time is given in units of the grid spacing $\Delta x$. For comparison
a dashed line $\propto 1/t^2$ is shown. After some initial
oscillations, for $N>3$ the scaling is very clean until; $t\sim 80$,
where finite size effects can become important.
clean until
\vspace{1cm}\\
{\bf Fig.~2}\\
The spectrum of the electric part of the Weyl tensor as a function of
$kt$ at time $t=8$ for a grid of size 160. On large 
scales, $kt/(2\pi)<1$, the spectrum is flat, white noise.
\vspace{1cm}\\
{\bf Fig.~3}\\The quantities $|(T_0^\mu;_\mu)|^2$ (dotted line), 
$|(T_i^\mu;_\mu)|^2$ (dashed lines) and $|( T_0^0/t)|^2$ (solid line)
are shown  as functions of $k$. The inaccuracy in energy
and momentum conservation is below 10\% for $k\le 32=128/4$. This lets
us that our code is accurate to better than 10\% for wavelengths of
 4 grid spacings and larger.
\vspace{1cm}\\
{\bf Fig.~4}\\The dashed curves and the triangles show 
$\dot{\beta}^2$ as a function of $t$ for fixed values of $k=n/n_{tot}$ 
for $n_{tot}=128$. The solid lines show the fits according
to the fitting formula given in the text.
\vspace{1cm}\\
{\bf Fig.~5}\\The crosses and triangles show $\dot{\beta}^2$ as a 
function of $n$, with $k=n/n_{tot}$ for fixed values of $t$. 
The solid curves show the fits according
to the fitting formula given in the text.
\vspace{1cm}\\
{\bf Fig.~6}\\ The dark matter contribution to the $C_\ell$'s from
a $(160)^3$ simulation. $(\ell+2)(\ell+1)\ell C_\ell/24$ is shown. 
For $\ell<20$ which is the dynamic range of this
simulation,  a white noise, $n=0$, spectrum fits reasonably well.
\vspace{1cm}\\
{\bf Fig.~7}\\The pixel distribution of $\de T/T$ for one
observer. 
\vspace{1cm}\\
{\bf Fig.~8}\\The values  $\ell(\ell+1)C_\ell/6$ for 27 observers are
plotted for $\ep=1$. 
The crosses are the individual observers and the solid line 
indicates the average. The sharp drop after $\ell\sim 30$ is due to
finite resolution (our dynamical range is approximately 25).
\vspace{1cm}\\
{\bf Fig.~9}\\
The $\chi^2$ contour plots for 66\%, 95\% and 99\%
confidence levels from fitting the $C_\ell,$ to a spectrum with
index $n$ with quadrupole amplitude $Q$ according to Eq.~(\ref{spec})
for $\ell \le 20$. In total 81 observers from 3 different $(192)^3$
simulations have been taken into account.
\vspace{1cm}\\
{\bf Fig.~10}\\ The root mean square of the temperature fluctuation at 
given angular scale is shown as a function of angle.
\vspace{1cm}\\
{\bf Fig.~11}\\The dark matter power spectrum (without bias and
nonlinear evolution). 
The result is averaged over 15  simulations on a $(256)^3$ grid
of two different physical scales. The error bar indicates one standard 
deviation.
\vspace{1cm}\\
{\bf Fig.~12}\\The dark matter pixel distribution from linear
perturbation theory. The positive skewness (0.76) and positive 
kurtosis (1.2) are clearly visible.

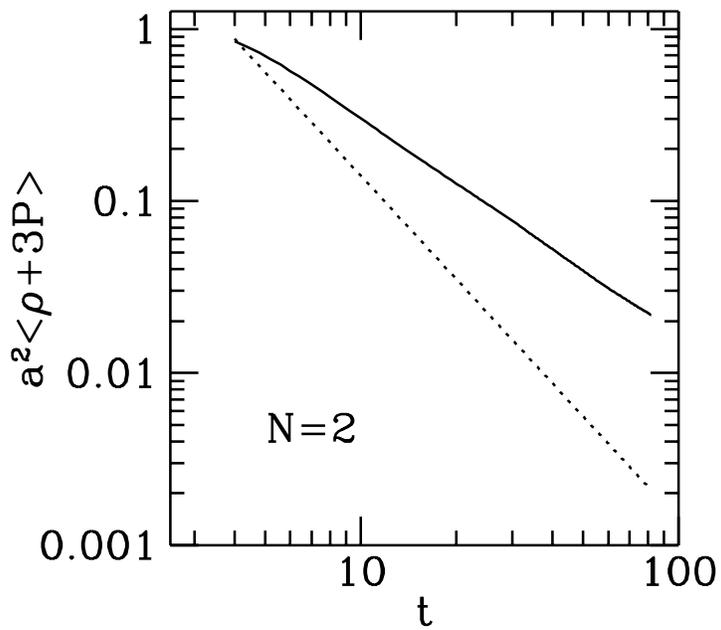
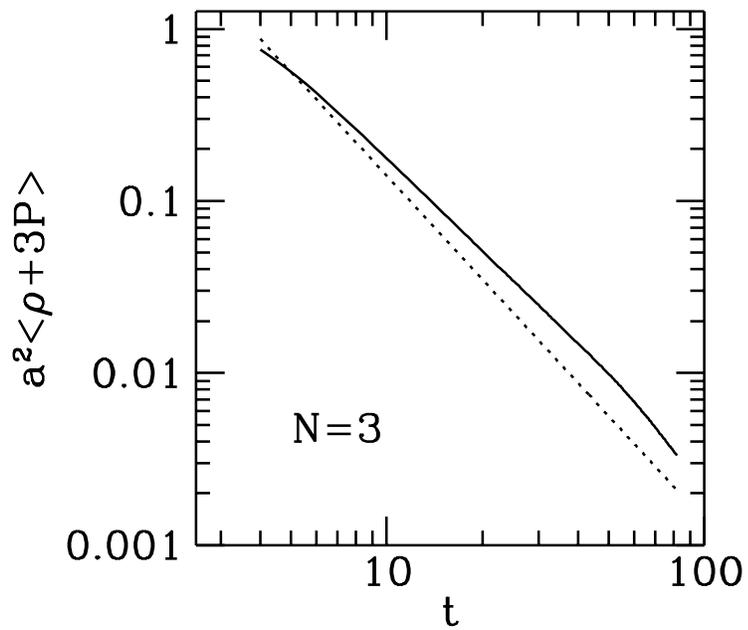
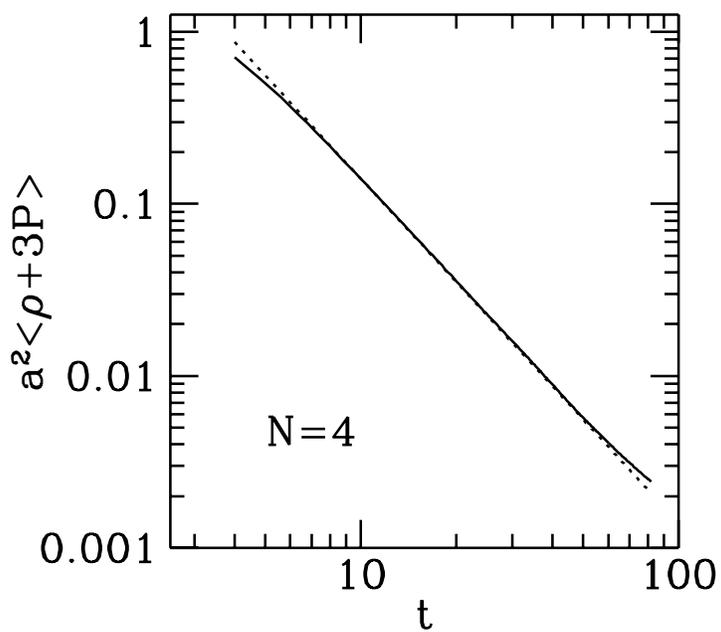
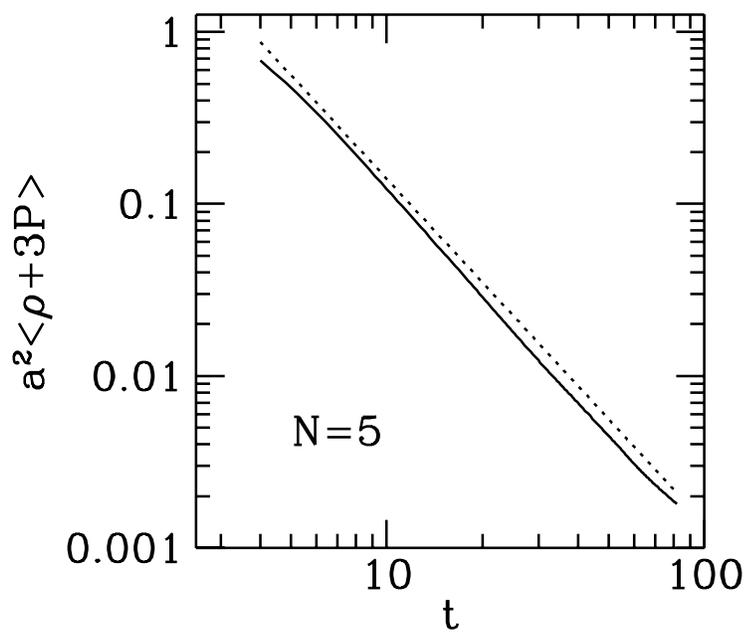
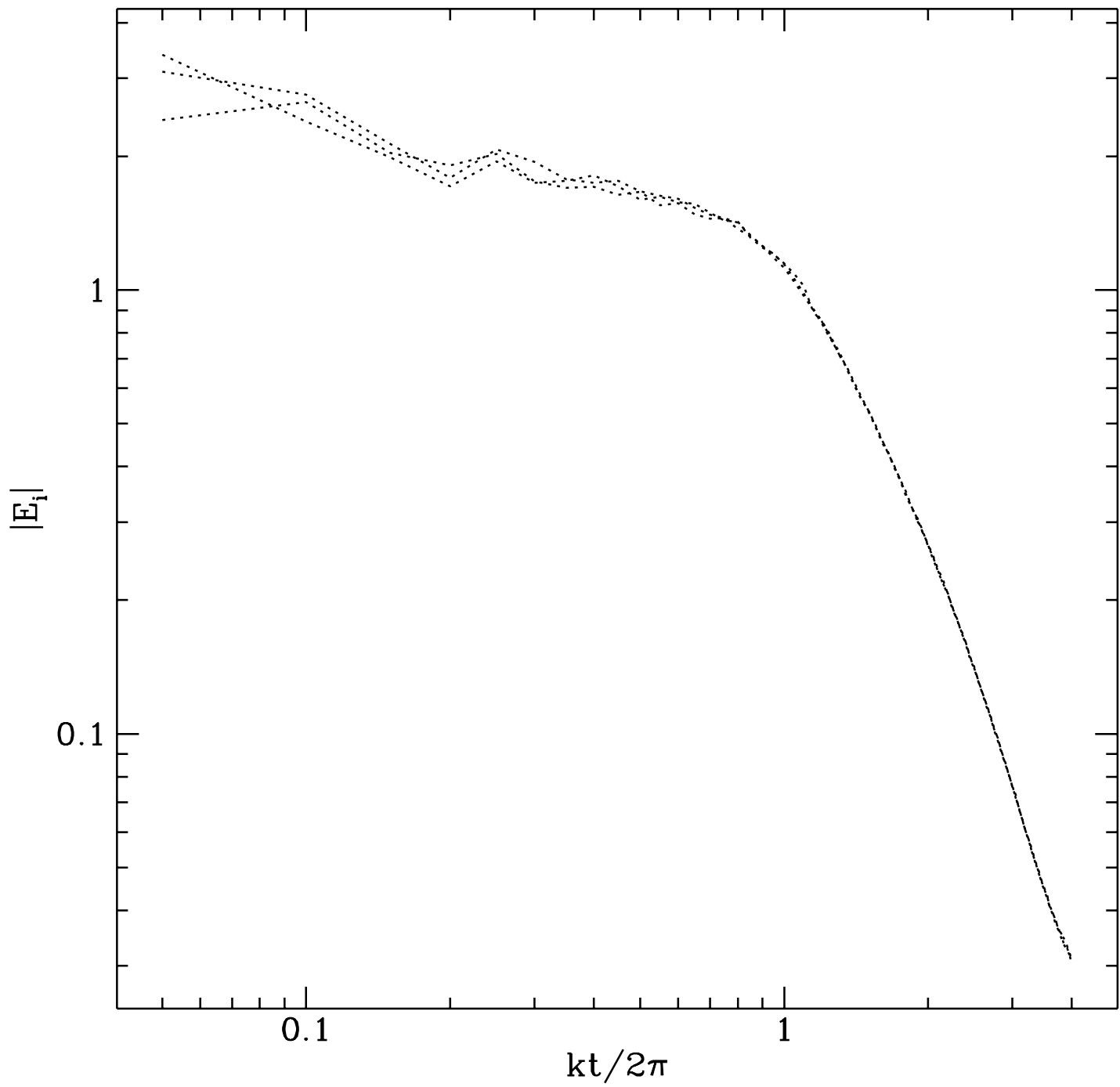
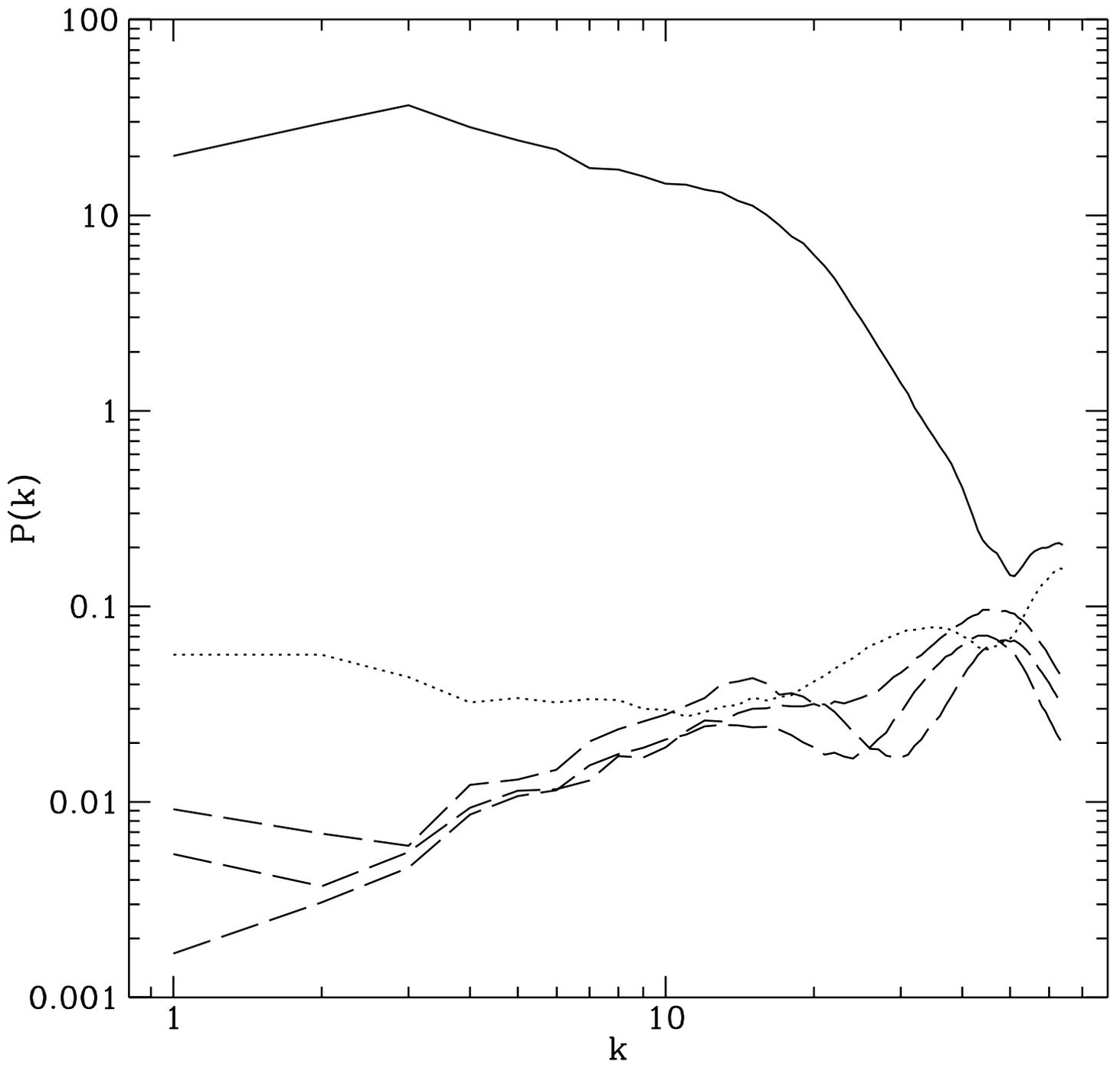
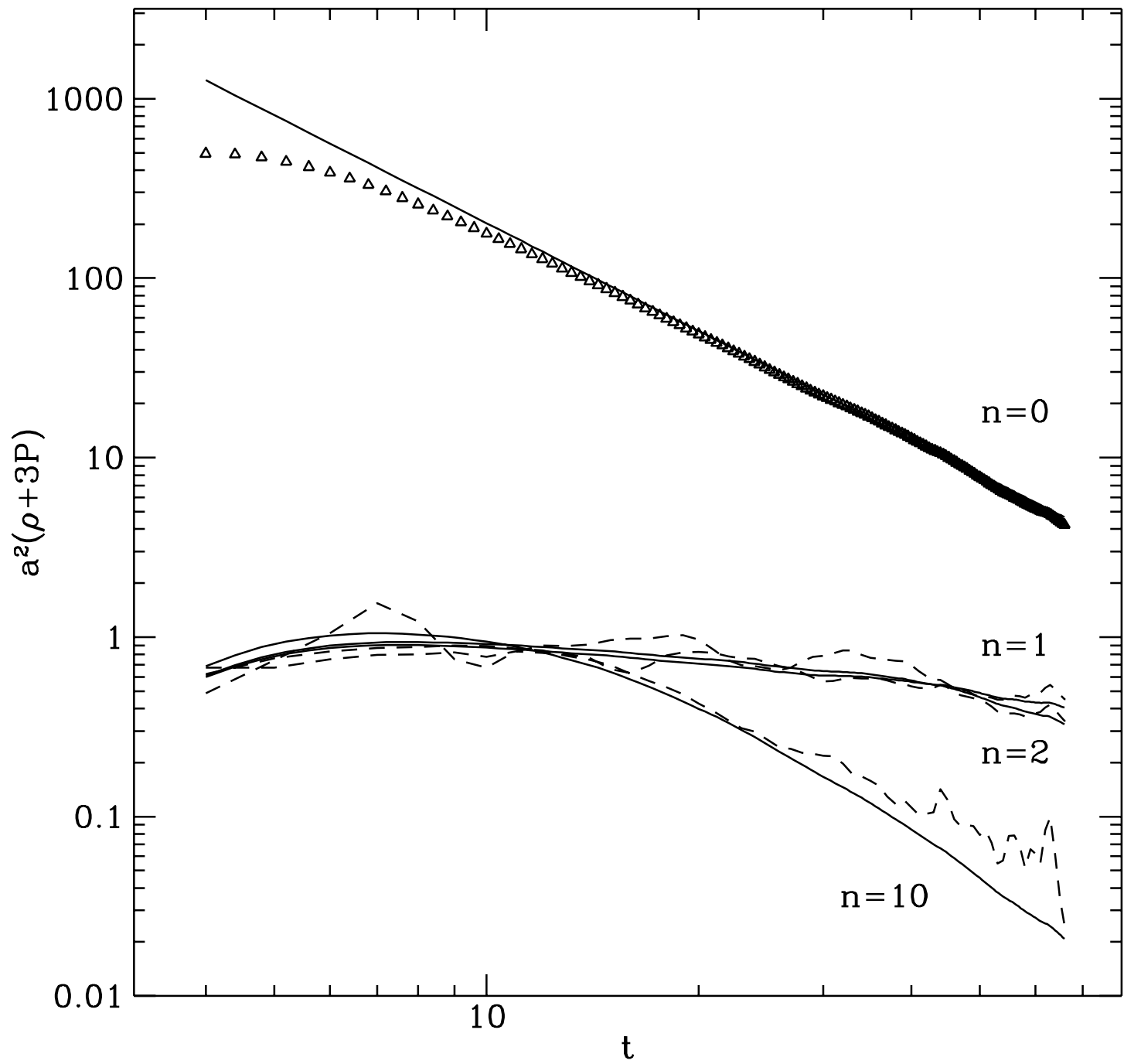
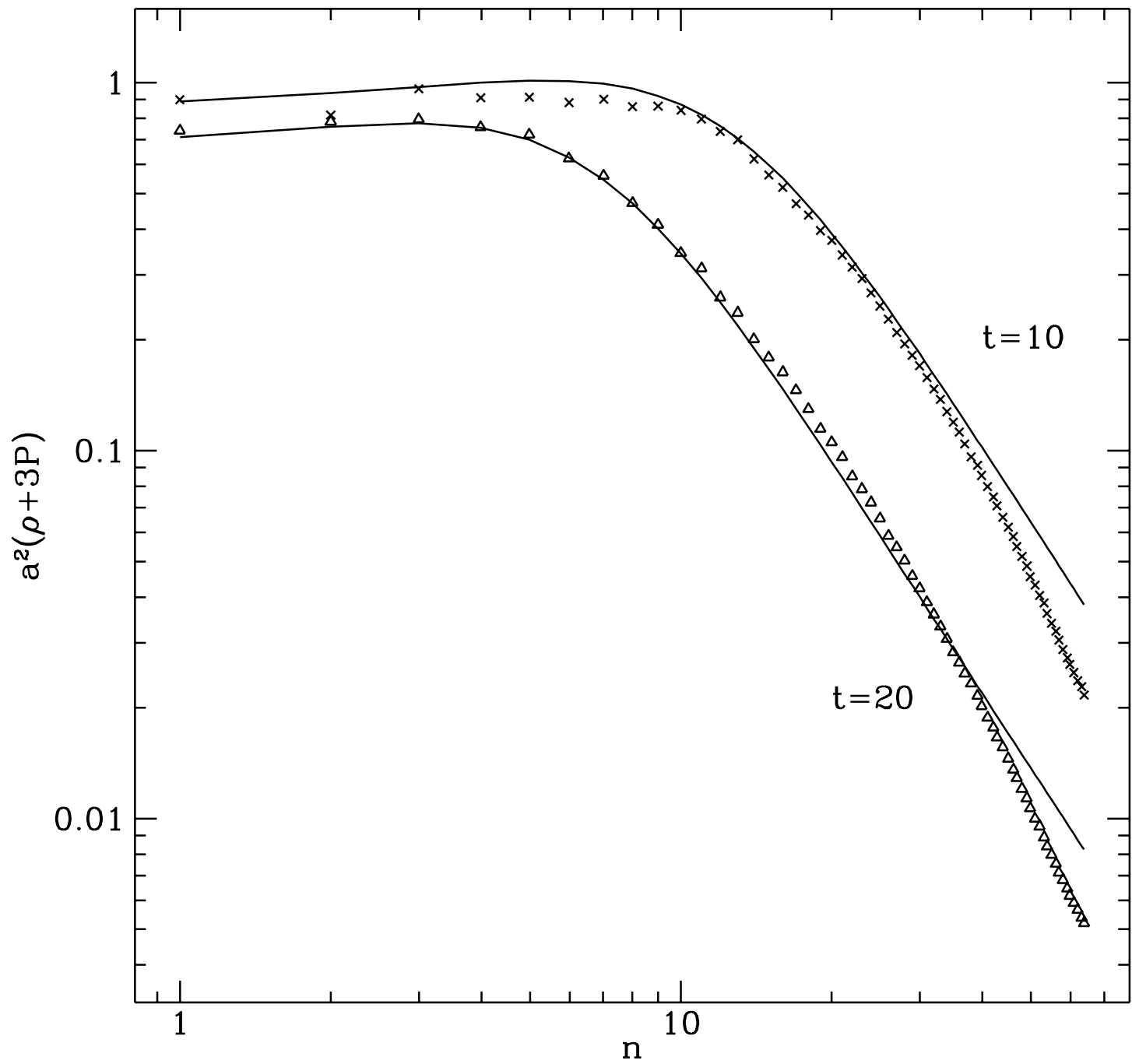
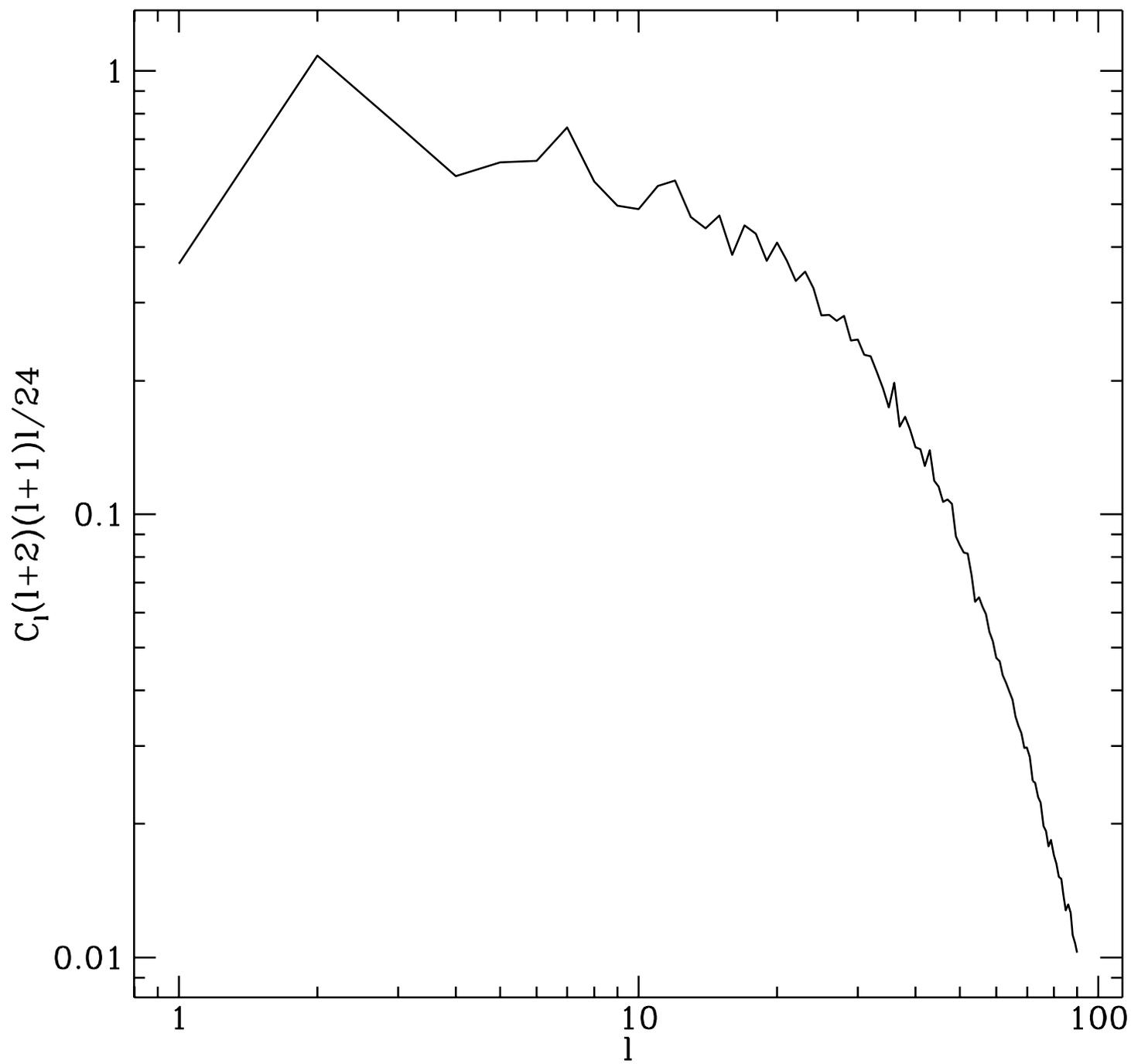
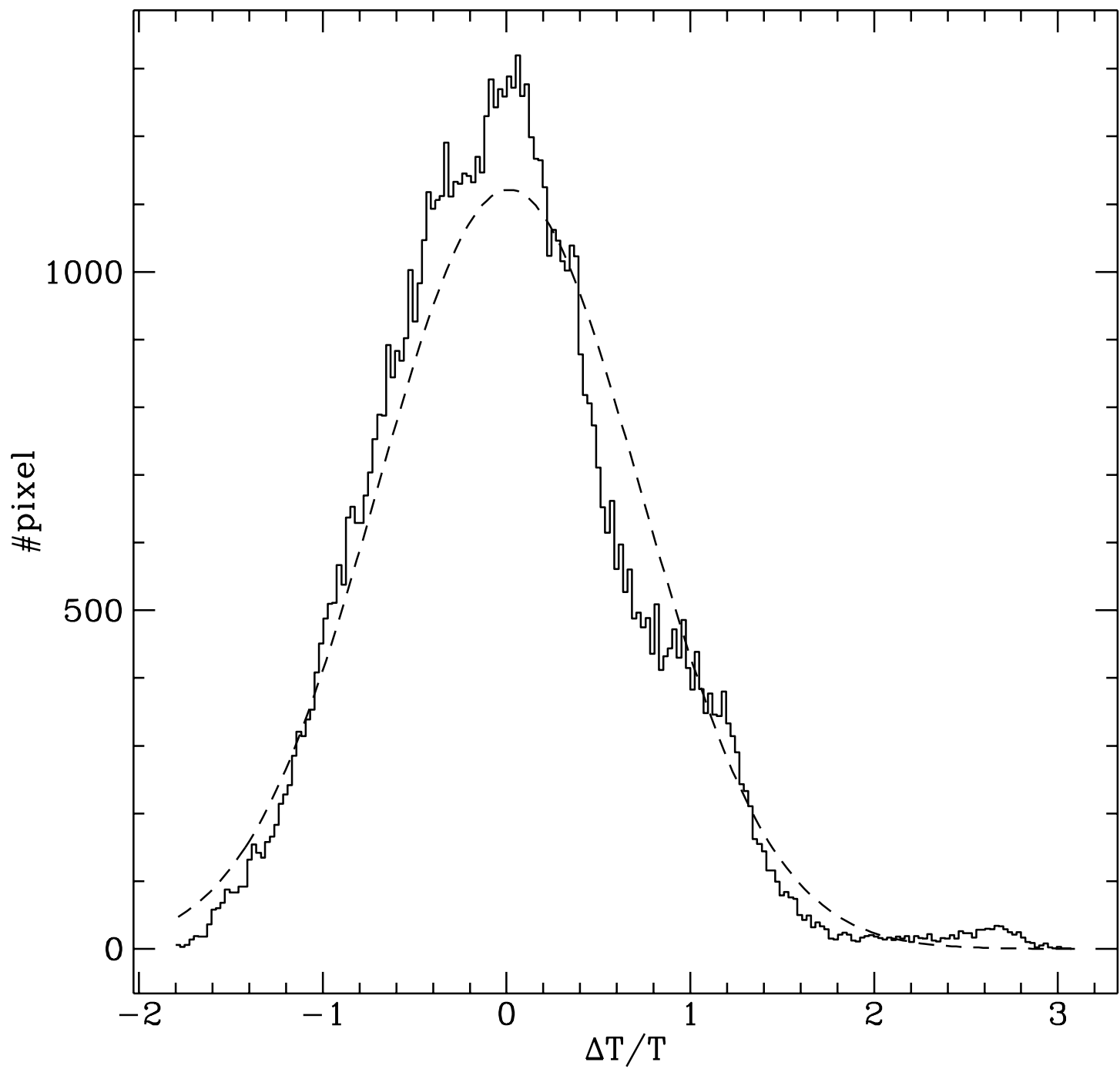
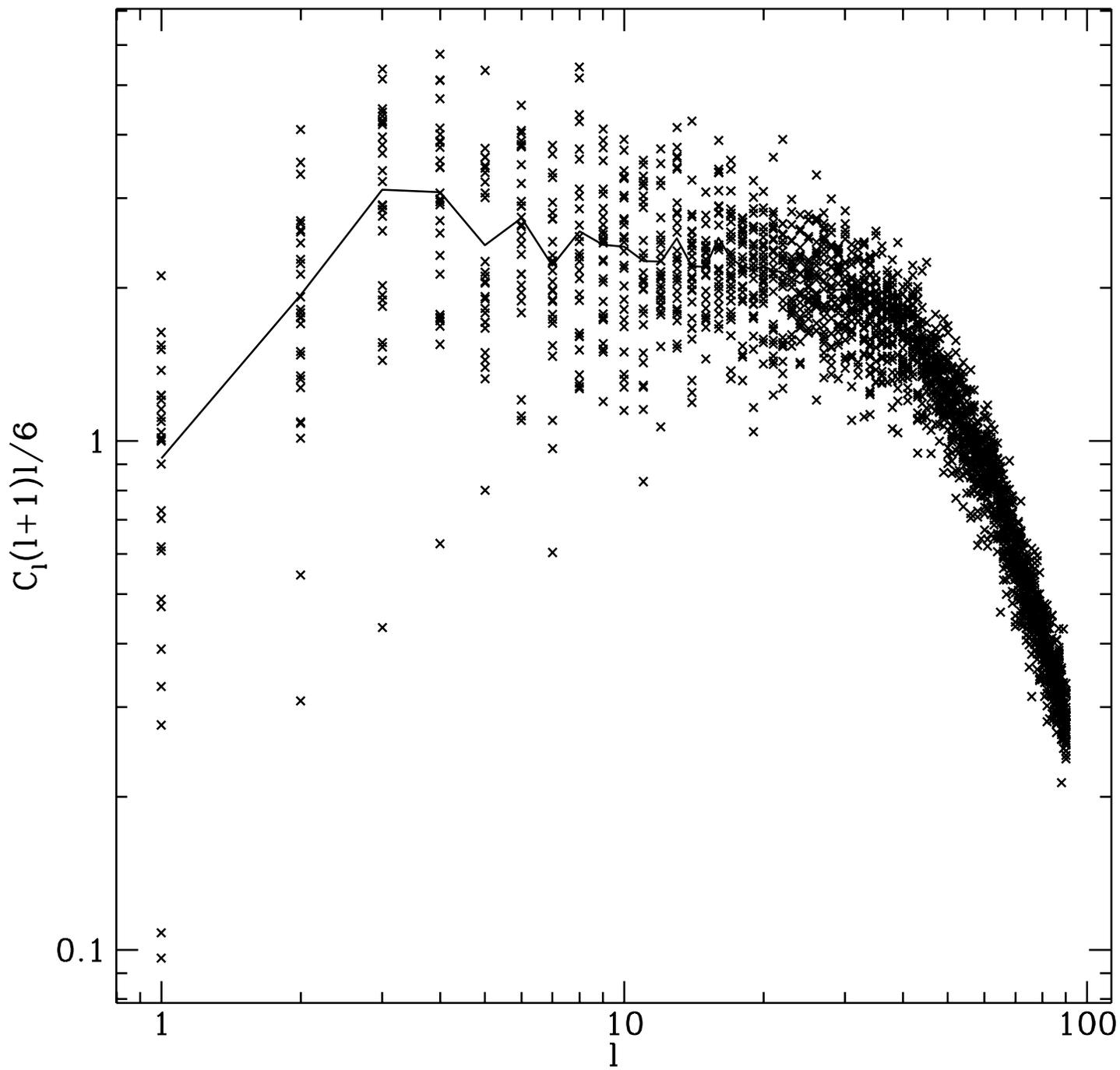
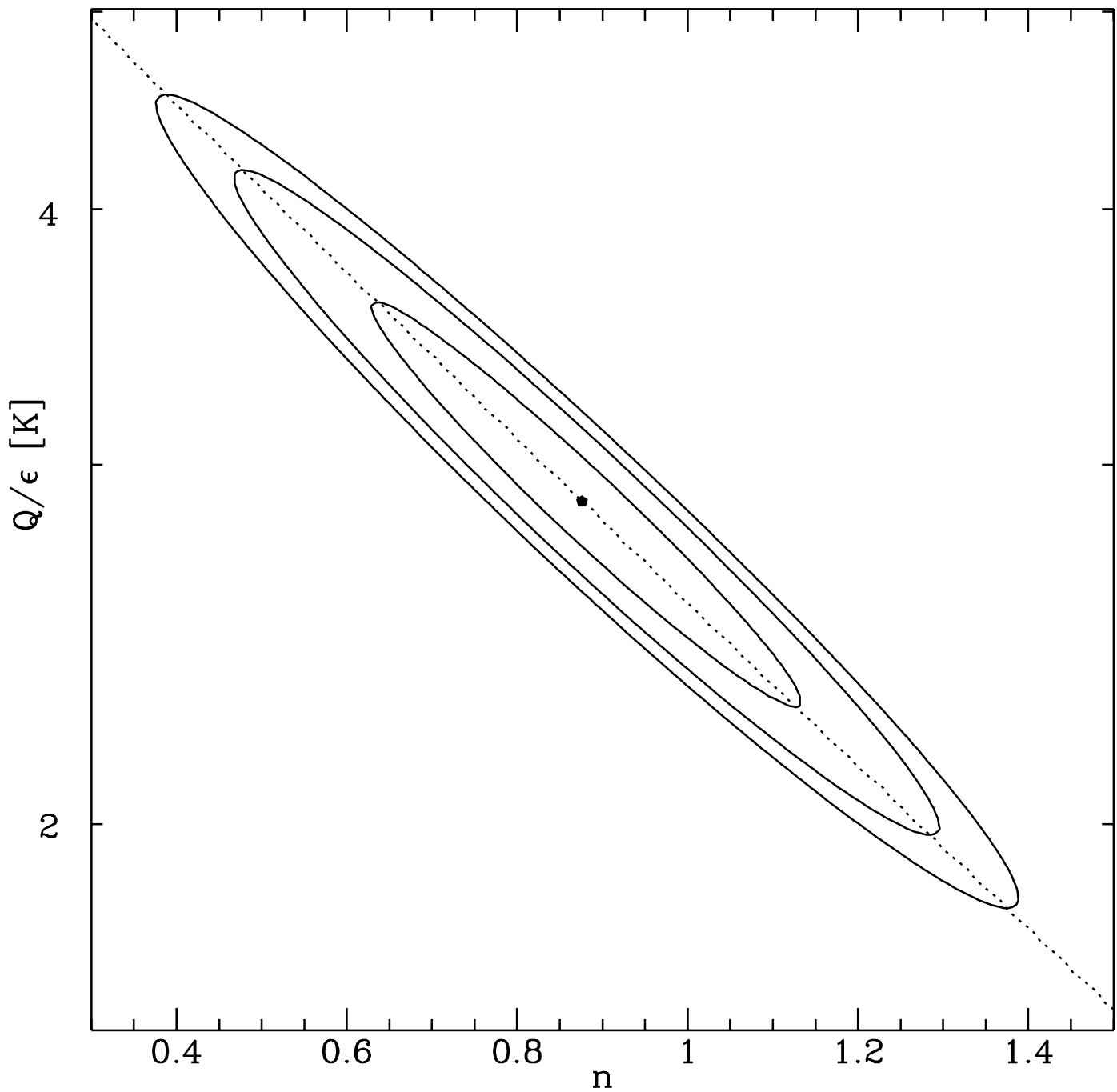
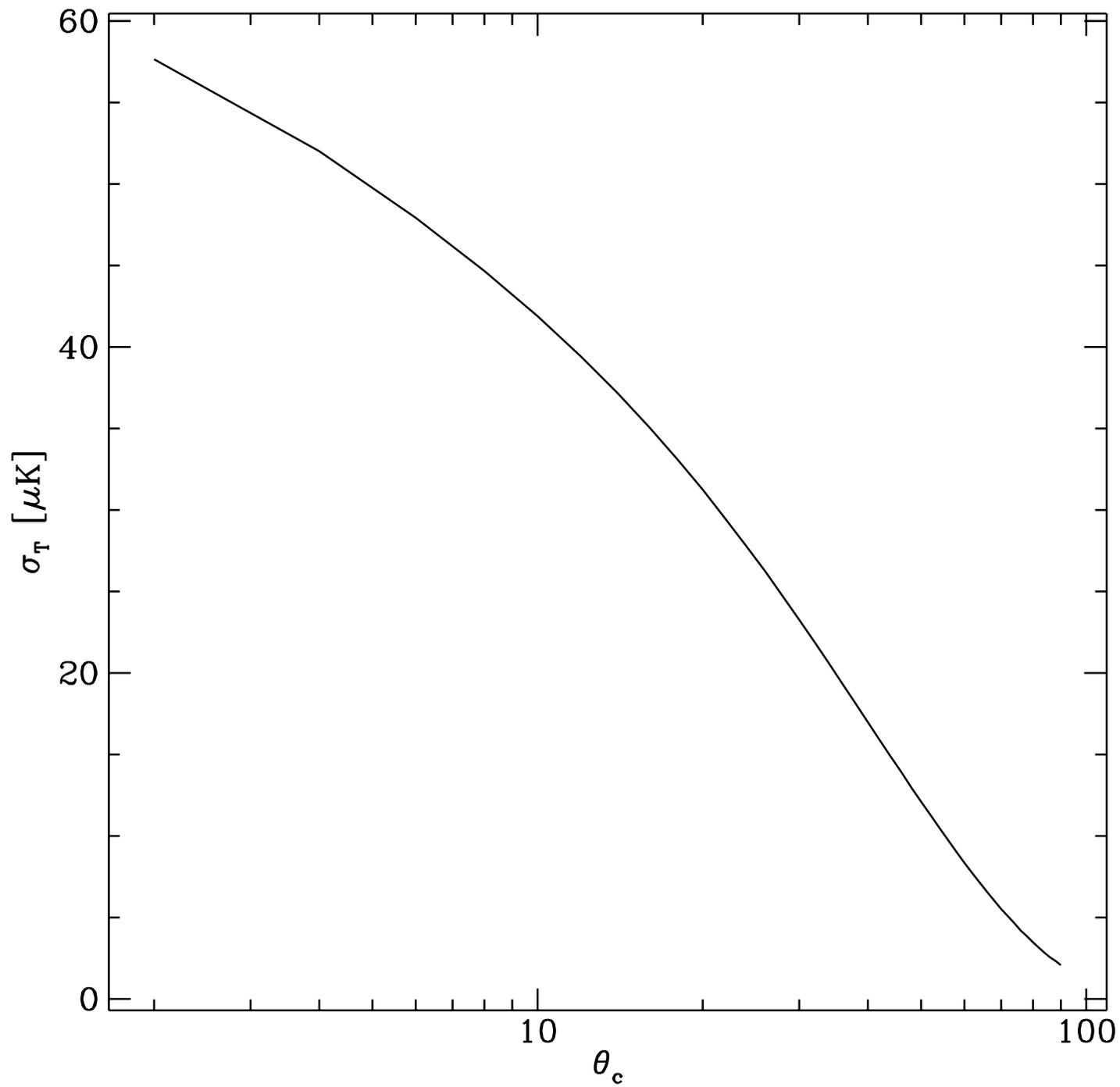
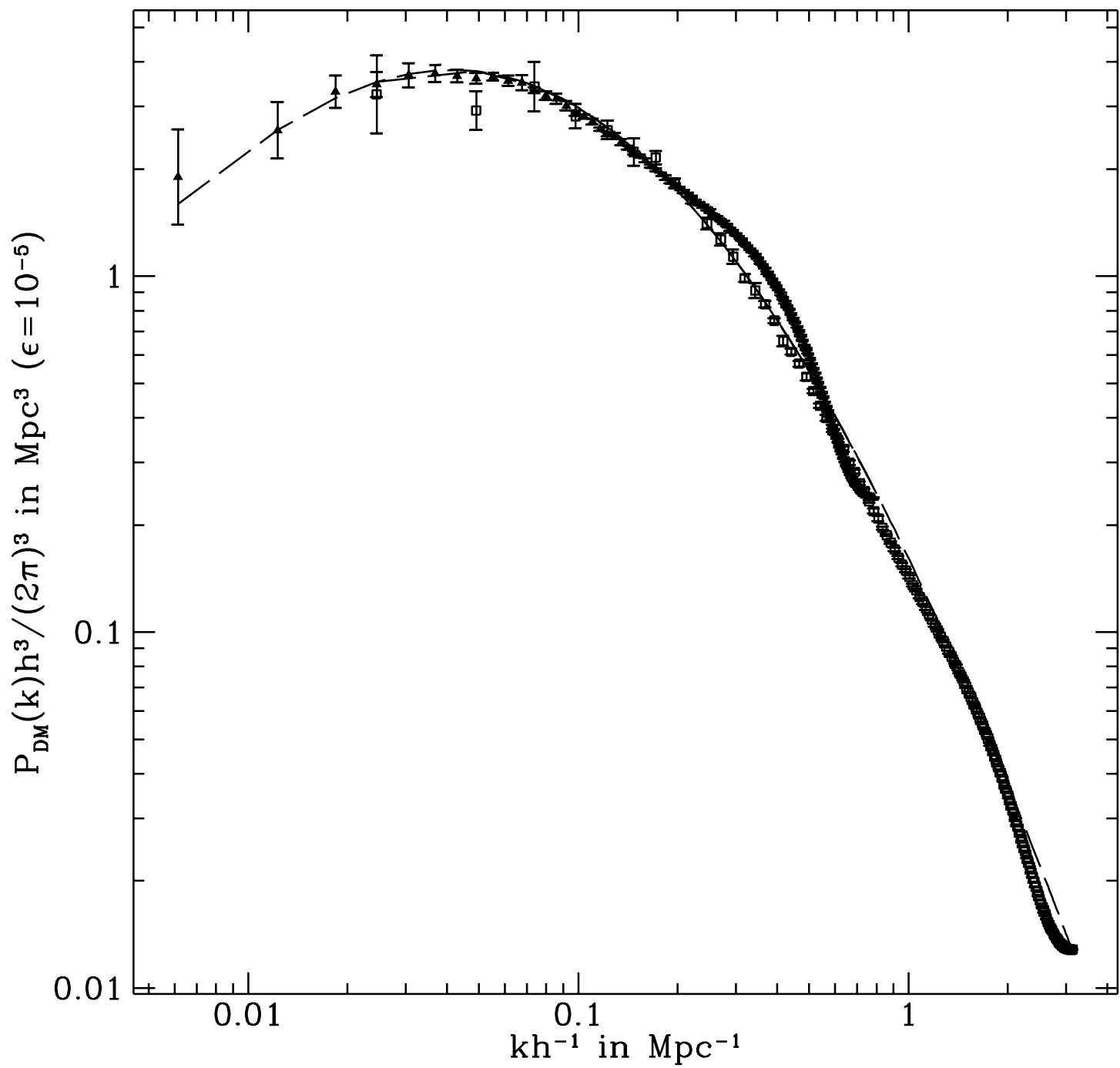
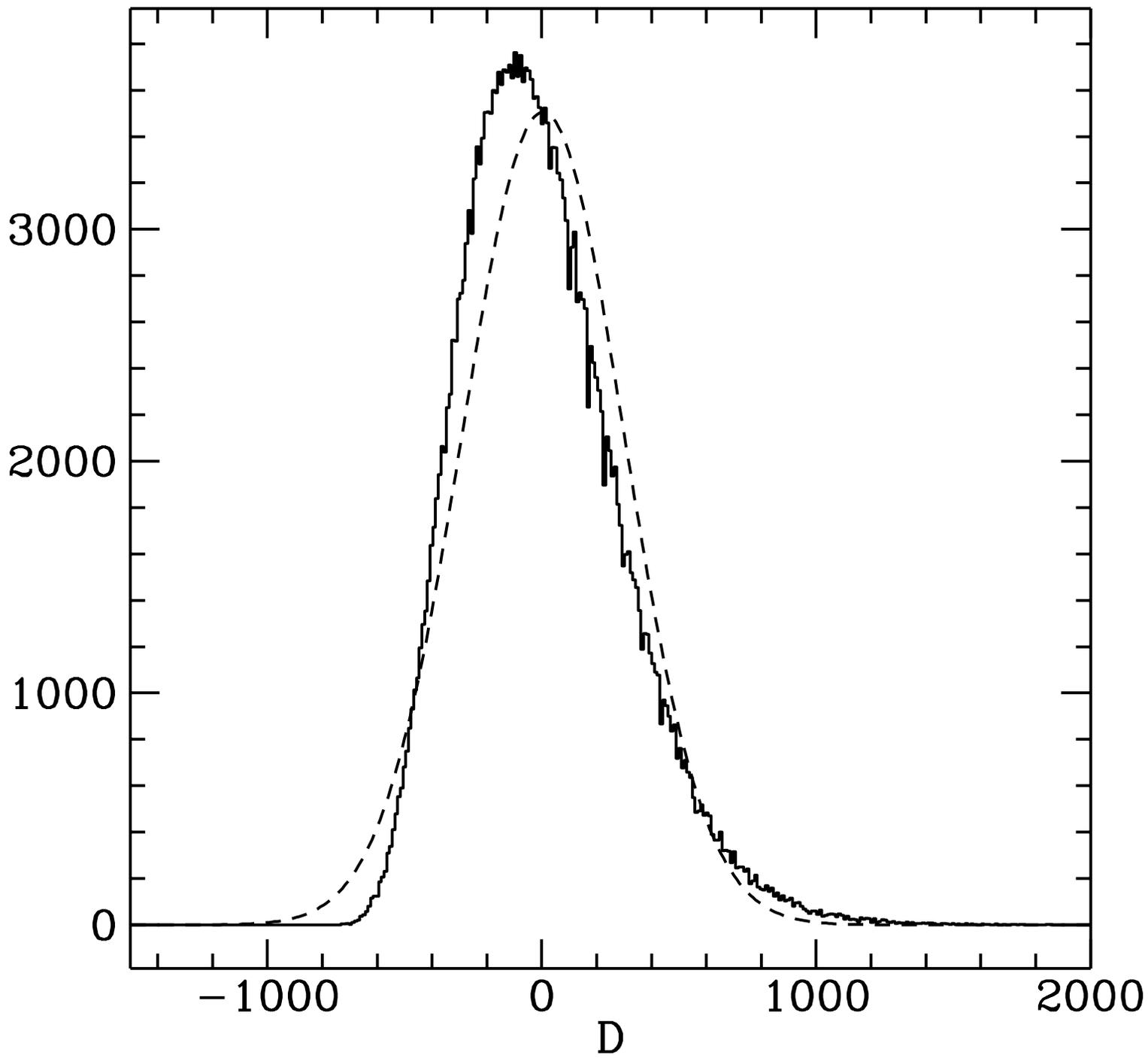
\end{document}